\documentclass[10pt,aps,pre,amsmath,amsfonts,amssymb,floatfix,longbibliography,superscriptaddress,notitlepage, onecolumn]{revtex4-1}
\usepackage{mathrsfs} \usepackage{graphicx,color} \usepackage{bbm} \usepackage{multirow}
\usepackage[normalem]{ulem}
\usepackage{lineno}

\newcommand\HY[1]{\textcolor{black}{#1}}

\begin{document}

 
\title{Computer simulations of the Gardner transition in structural glasses}

\author{Yuliang Jin}
\affiliation{CAS Key Laboratory of Theoretical Physics, Institute of Theoretical Physics, Chinese Academy of Sciences, Beijing 100190, China}
\affiliation{School of Physical Sciences, University of Chinese Academy of Sciences, Beijing 100049, China}
\affiliation{Wenzhou Institute, University of Chinese Academy of Sciences, Wenzhou, Zhejiang 325000, China}

\author{Hajime Yoshino}
\affiliation{Cybermedia Center, Osaka University, Toyonaka, Osaka 560-0043, Japan}
\affiliation{Graduate School of Science, Osaka University, Toyonaka, Osaka 560-0043, Japan}

\begin{abstract}
\end{abstract}

\maketitle
\tableofcontents

\section{{Connections between Gardner and spin-glass transitions}}
\label{sec-connection-to-SG}

The exact mean-field theory 
for the simplest glass-forming system - the dense assembly of hard spheres 
in the large dimensional limit
- predicts the existence of a Gardner phase~\cite{parisi2020theory,berthier2019gardner}. 
This transition is characterized by full replica symmetry breaking (RSB) that implies two fascinating physical consequences. 
(i) A
hierarchical free-energy landscape, i.e.,   the thermal fluctuations are organized hierarchically, 
meaning that configurations are grouped into meta-basins that are further grouped into meta-meta basins, ... (ii) The marginal stability, i.e., the system responds sensitively 
to infinitesimal perturbations. 
Here we discuss recent results of numerical simulations to examine these mean-field predictions in physical dimensions. 

\begin{figure}[h]
\centerline{\includegraphics[width=0.8\textwidth]{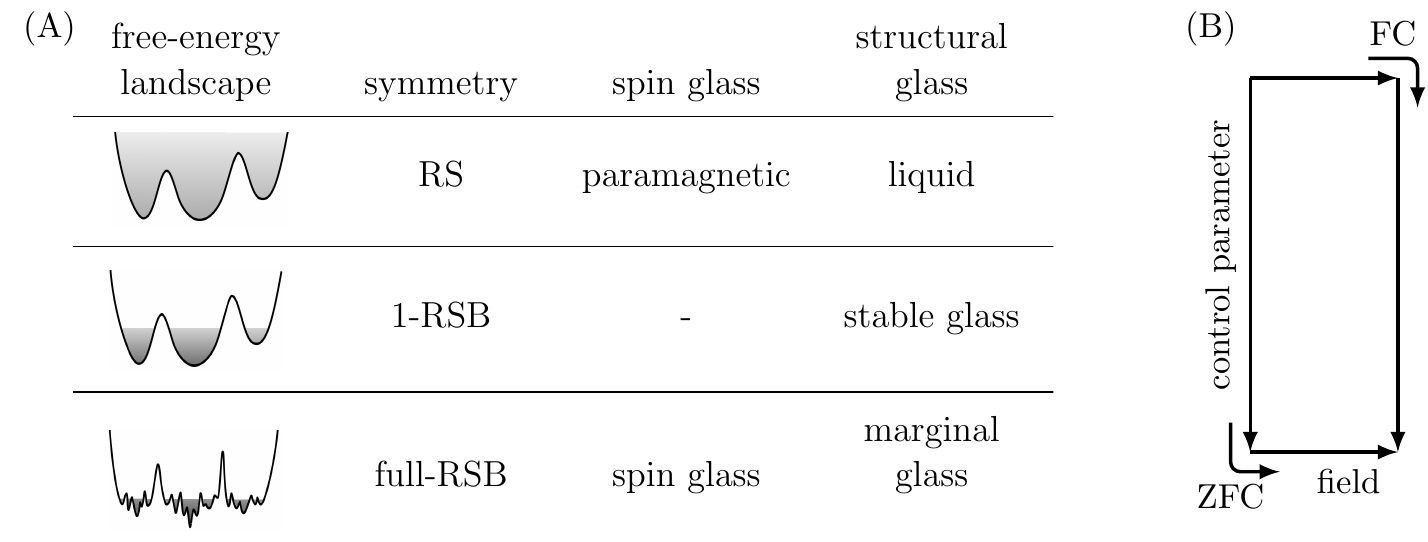}}
\caption{
{\color{black} (A) Correspondence between spin and structural glasses. Note that one step RSB (1-RSB) exists in certain spin-glass models such as the spherical p-spin ($p>2$) model~\cite{castellani2005spin}.
(B) Schematic of ZFC/FC protocols. 
In spin and hard-sphere  glasses, the control parameter is the temperature $T$ and the density (volume fraction) $\varphi$ respectively, and the external field is the magnetic field $h$ and the shear-strain $\gamma$ respectively.}
}
\label{fig:landscape}
\end{figure}

From the viewpoint of RSB, the Gardner transition in structural glasses belongs to the same full RSB universality class of the spin-glass transition (see Fig.~\ref{fig:landscape}(A)). 
This theoretical ground motivates us to borrow ideas from the extensive research on spin-glasses to study the Gardner transition. 
To this end, it is useful to review firstly some of the essential results obtained in 
spin-glass experiments and simulations. 

\begin{figure}[h]
\centerline{\includegraphics[width=0.8\textwidth]{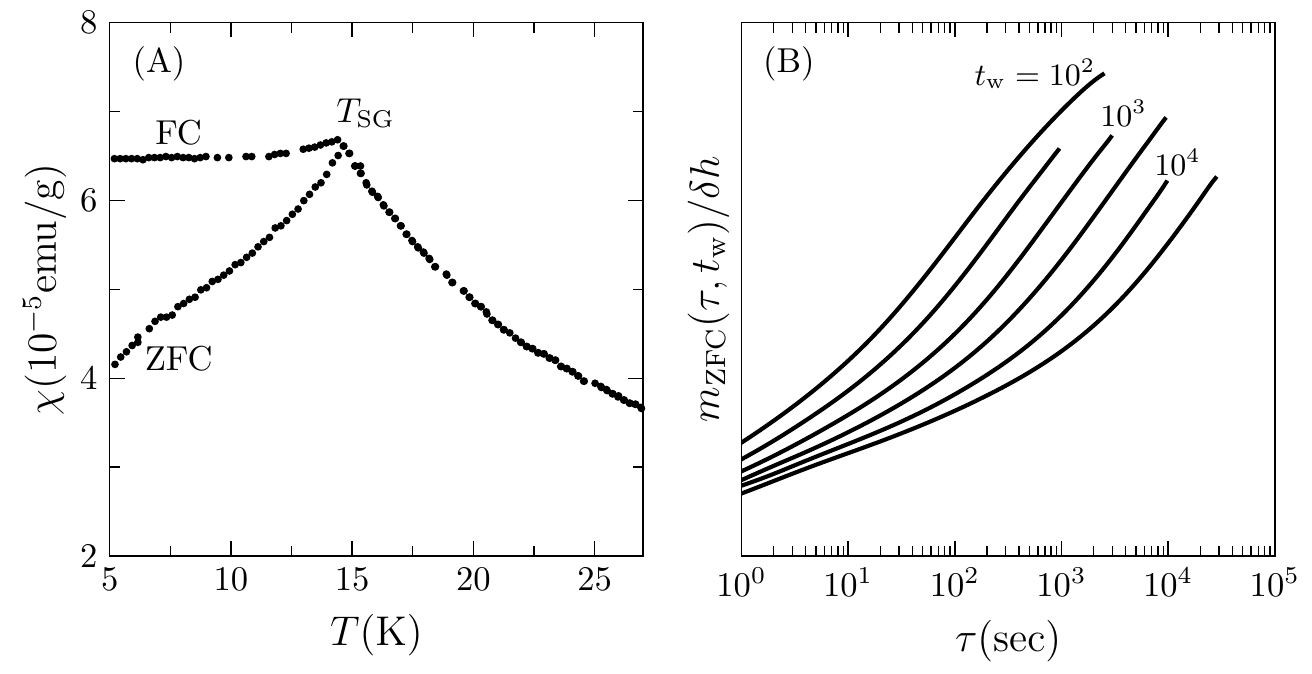}}
\caption{Experimental results on CuMn spin-glasses. 
(A) ZFC/FC susceptibilities (adapted from \cite{nagata1979low}), and (B) the time evolution of $m_{\rm ZFC}(\tau, t_{\rm w})/\delta h$ in an aging experiment (adapted from \cite{granberg1988observation}).
}
\label{fig:SG}
\end{figure}

 The RSB solution  immediately implies a hierarchy of linear responses through the fluctuation-dissipation relation~\cite{MPV87}. One expects short-time, intermediate-time, and long-time linear responses associated with thermal fluctuations inside basins, meta-basins, and meta-meta-basins.
A remarkable consequence  is the ``anomaly" 
\HY{that gives a natural explanation for the}
protocol-dependent linear responses observed experimentally ~\cite{nagata1979low}.
In one protocol called  {\it field cooling} (FC), one measures the magnetization $m_{\rm FC}$ of a spin-glass under cooling from a high temperature $T_{\rm max}$ down to a low temperature $T_{\rm min}$ below the spin-glass transition temperature $T_{\rm SG}$ in the presence of a weak external magnetic field $\delta h$; in the other protocol called {\it zero field cooling} (ZFC), one cools  the spin-glass from $T_{\rm max}$ down to $T_{\rm min}$ without the field ($h=0$), then switches on the magnetic field $\delta h$ and measures the magnetization $m_{\rm ZFC}$ under heating the spin-glass back to $T_{\rm max}$ {\color{black}(see Fig.~\ref{fig:landscape}(B))}. The two susceptibilities $\chi_{\rm FC}=m_{\rm FC}/\delta h$ and $\chi_{\rm ZFC}=m_{\rm ZFC}/\delta h$ are the same above $T_{\rm SG}$, but different ($\chi_{\rm FC} > \chi_{\rm ZFC}$) below {\color{black}(see Fig.~\ref{fig:SG}(A))}. 
The fact that $\chi_{\rm FC}$ and $\chi_{\rm ZFC}$ are different in the spin-glass phase is referred to as an ``anomaly", because the susceptibility is protocol-independent in standard magnetic systems. 
The RSB theory gives
$\chi_{\rm FC} - \chi_{\rm ZFC} = \beta \left [ \int_{0}^{1} dq P(q)q-q_{\rm EA} \right]$, with $\beta$ the inverse temperature. Here $q_{\rm EA}$ is the Edwards-Anderson (EA) order parameter~\cite{edwards1975theory} representing the strength of the thermal fluctuation within lowest-level basins, while $\int_{0}^{1} dq P(q)q$ represents the integral of  thermal fluctuations coming from all levels in the hierarchy. In the replica symmetric (RS) solution, $P(q)=\delta(q-q_{\rm EA})$, so that the anomaly vanishes. 

The anomaly is known to be not a transient but a long-time effect,  as demonstrated by a series of experiments that reveal aging effects in spin-glasses \cite{granberg1988observation,vincent1997slow,nordblad1998experiments}. 
To study the dynamical effects, one can generalize the ZFC protocol by introducing a waiting time $t_{\rm w}$ before switching on the magnetic field, and a measurement time $\tau$ elapsed in the presence of the field.
By increasing $\tau$,  $m_{\rm ZFC}(\tau,t_{\rm w})$ increases passing through $m_{\rm ZFC}$ and heads toward $m_{\rm FC}$ (see Fig.~\ref{fig:SG}(B)). However, $m_{\rm ZFC}(\tau,t_{\rm w})$ does not reach $m_{\rm FC}$ within finite time, and its  time evolution 
as a function of $\tau$ slows down with increasing waiting time $t_{\rm w}$, manifested by scaling laws depending on $\tau/t_{\rm w}$. 
These experimental observations are significant because they reveal the out-of-equilibrium nature of spin-glasses. 
To describe the aging effects and the anomaly from a purely dynamical point of view,
a dynamical mean-field theory on spin-glass models is developed \cite{cugliandolo1993analytical,cugliandolo1994out}.
The dynamical theory relates RSB to the notion of effective
temperature that characterizes out-of equilibrium glassy dynamics \cite{cugliandolo1997energy,franz1998measuring}.
\HY{The numerical evidence of  effective temperature 
\cite{marinari2000off}
and  non-zero anomaly in the long-time limit
\cite{yoshino2002extended}
has been indicated by detailed  simulations
of finite-dimensional spin-glass models.}


The marginal stability of the spin-glass phase \HY{may account for}
various complex non-linear responses,
such as the effect of static chaos  with respect to an infinitesimal change of  temperature,
or avalanches with respect to an infinitesimal change of magnetic field.
Indeed the equilibrium spin configurations at large length scales are completely reshuffled
by infinitesimal perturbations, which is
predicted first by the droplet theory \cite{bray1987chaotic,fisher1988equilibrium,fisher1988nonequilibrium},
and later by theories based on RSB \cite{kondor1989chaos,rizzo2003chaos,rizzo2006chaos,yoshino2008stepwise,parisi2010chaos,le2010avalanches,le2012equilibrium,franz2017mean}. The rejuvenation-memory effects observed experimentally
\cite{jonason1998memory}
may be related to such non-linear responses ~\cite{yoshino2001multiple,jonsson2004nonequilibrium}.

Once one is aware of the correspondence between spin and structural glasses (see Fig.~\ref{fig:landscape}(A)), it is natural to use strategies inherited from spin-glass studies to explore the physics of Gardner phase in structural glasses. For example, in the ZFC/FC protocols, the role of the magnetic field $h$ for spin-glasses can be replaced by the shear strain $\gamma$ for structural glasses (see Fig.~\ref{fig:landscape}(B)), which only changes the boundary condition but not the thermodynamic properties of the bulk. Indeed, the replica theory of structural glasses 
predicts
a
hierarchy of shear moduli reflecting RSB~\cite{yoshino2010emergence,YO12,yoshino2014shear}. 
By adapting the methods developed in spin-glasses, one can examine the aging effects, the protocol-dependent linear responses, and the non-linear responses such as avalanches,  in structural glasses with respect to shear deformations. 
Our discussion focuses on one of the simplest models of structural glasses in three dimensions, hard spheres, where 
 the (reduced) pressure $p$ (or the volume fraction $\varphi$) plays the role of temperature $T$.
According to the replica theory~\cite{parisi2020theory} \HY{that is exact in the large
dimensional limit},
a Gardner transition occurs in hard spheres under both compression and shear,  which is examined by simulations 
\HY{at three-dimensions} in the following sections. 
 \HY{The dynamical mean-field theory for the hard-sphere glass has also been set up \cite{maimbourg2016solution,kurchan2016statics,agoritsas2019outA,agoritsas2019outB}, but a detailed theoretical analysis of the out-of equilibrium dynamics remains challenging. Nonetheless, the analogy to the spin-glass problem outlined above allows us to infer the implications of RSB on the dynamics of hard spheres.}

\section{Gardner transition under compression}

\subsection{Preparation of ultra-stable glasses}

To study the Gardner transition, 
we must prepare a glass at first.
Experimentally, glasses are obtained by
a slow thermal or compression annealing, the rate of which
determines the location of the glass transition. 
It is found that
a detailed numerical analysis of the Gardner transition requires
the preparation of extremely well-relaxed glasses
(corresponding to structural relaxation timescales challenging to simulate in standard algorithms), in order to study vibrational motions of particles  without interference from 
 diffusion. 
Such ultra-stable glasses can be numerically generated by applying a swap Monte-Carlo scheme~\cite{kranendonk1991computer, grigera2001fast} to a  
 simple glass-forming model -- 
a polydisperse mixture of {\color{black} $N$ hard spheres~\cite{berthier2016growing}}.


The annealing procedure contains two steps~\cite{berthier2016growing}.
First, one produces equilibrated liquid configurations
at various densities
$\varphi_{\rm g}$ with the help of the swap algorithm.
Second, starting from these liquid configurations,
one switches to 
standard molecular dynamics simulations~\cite{lubachevsky1990geometric} during which
the system is compressed out of equilibrium up to 
target  densities $\varphi > \varphi_{\rm g}$.
In order to obtain thermal and disorder averaging, this
procedure is repeated over many samples, 
each corresponding to different initial equilibrium configurations at $\varphi_{\rm g}$, and over many 
independent quench
realizations for each sample. The independent realizations of the same sample have identical particle positions at $\varphi_{\rm g}$, but are assigned to different initial velocities drawn from the Maxwell–Boltzmann distribution.

The above numerical protocol is analogous to thermal annealing with different 
 cooling rates, which results in  different glass transition temperatures.
Each glass transition density $\varphi_{\rm g}$ selects a particular
glass state. The value of $\varphi_{\rm g}$ ranges from  the mode-coupling theory (MCT) density (or the dynamical glass transition density) $\varphi_{\rm d}$, at which the liquid relaxation is slow but affected by activated $\alpha$-processes,  to $\varphi_{\rm g} \gg \varphi_{\rm d}$, where 
particle diffusion and vibrations  are fully separated. 
For sufficiently large $\varphi_{\rm g}$,
the $\alpha$-relaxation time becomes larger than the simulation time by many orders of magnitude;
one thus
obtains unimpeded access to the dynamics within the glass state, i.e., the $\beta$-relaxation 
processes~\cite{Goldstein2010}.

\begin{figure}[h]
\centerline{\includegraphics[width=0.5\textwidth]{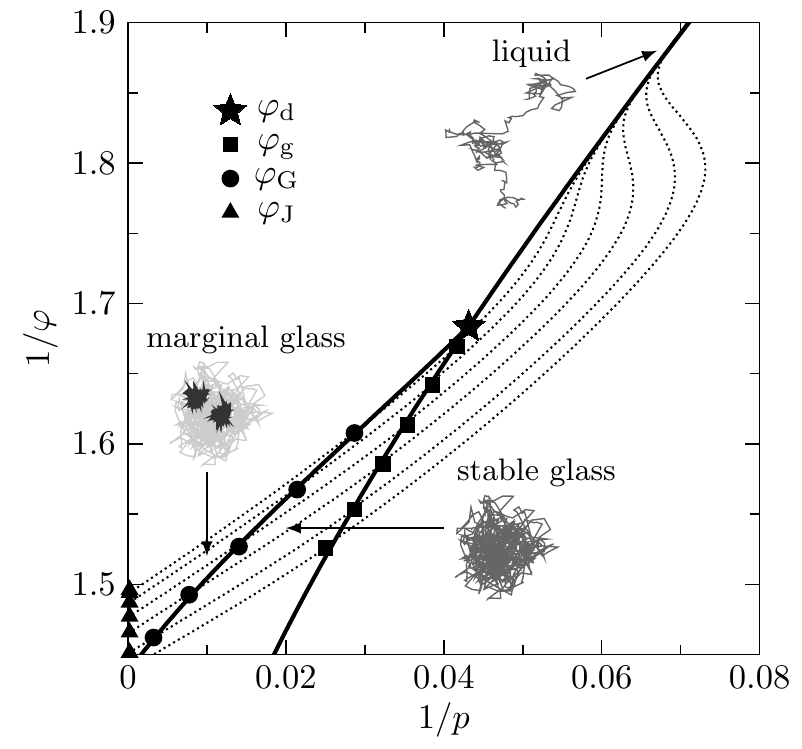}}
\caption{ Phase diagram of a polydisperse hard-sphere glass in three dimensions (adapted from~\cite{berthier2016growing}). Solids lines represent the CS liquid EOS and the Gardner line, and dashed lines represent glass EOSs. The insets show typical particle motions in three phases. 
}
\label{fig:illustration_with_PD}
\end{figure}

The liquid equation of state (EOS) for the reduced pressure $p=\beta P/\rho$ of the model, 
where $\rho$ is
the number density,  
and $P$ the system pressure, can be well described by  the Carnahan-Starling (CS) equation~\cite{B70}.  
The dynamical glass transition density $\varphi_{\rm d} = 0.594(1)$ was estimated following
 the strategy in Ref.~\cite{Charbonneau2014}. Note that the dynamical glass transition is only rigorous in large dimensions; it becomes a dynamical crossover in three dimensions {\color{black} (see Chapter~16 for a detailed discussion}).
The non-equilibrium glass EOSs
associated with compression  terminate 
 at inherent states (where $p \to \infty$) that correspond to, for hard spheres, 
jammed configurations at $\varphi_{\rm J}$, and
can be captured by a free-volume scaling form, $p_{\rm glass} (\varphi ) \sim (\varphi_{\rm J} - \varphi)${\color{black}~\cite{donev2005pair}}.  {\color{black} Figure~\ref{fig:illustration_with_PD}} presents the  phase diagram and EOSs  of the model.

Along each glass EOS of a given $\varphi_{\rm g}$, \HY{a corresponding Gardner transition may exist at density $\varphi_{\rm G}$ (or pressure $p_{\rm G}$),  as predicted by the mean-field theory}.
For $\varphi_{\rm g} < \varphi < \varphi_{\rm G}$, the system is in a {\it stable glass} phase: each glass state is confined in one of the structureless basins on the free-energy landscape (see {\color{black} Fig.~\ref{fig:landscape}(A)}), and is stable in response to small mechanical deformations. On the other hand,  the regime $\varphi_{\rm G} < \varphi < \varphi_{\rm J}$ corresponds to a {\it marginal glass} phase, where each simple glass basin splits into a fractal hierarchy of sub-basins and the glass becomes marginally stable to deformations. The Gardner line and the liquid EOS merge around $\varphi_{\rm d}$, suggesting the mixing of dynamical behavior associated to the Gardner transition and to the glass transition   -- this is why one needs to focus on ultra-stable glasses in order to explore pure Gardner physics.


\subsection{Key observables and protocols}

In the glass state, particles vibrate inside  their cages (see the insets of {\color{black} Fig.~\ref{fig:illustration_with_PD}}).
The first approach to study the Gardner transition is based on the direct analysis of {\it caging order parameters}, which quantify the  caging properties of particles. 
In the case of stable glasses, the caging order parameter is defined as,
$\Delta_{\rm EA}=\lim_{t\to \infty}\frac{1}{N} \sum_{i=1}^{N} \langle|\vec{r}_i(t) - \vec{r}_i(0)|^2 \rangle$, where $\vec{r}_{i}(t)$ is the position of particle $i$ at time $t$.
The parameter $\Delta_{\rm EA}$, which decreases with the degree of annealing, corresponds to nothing but the EA  parameter $q_{\rm EA}$ in spin-glasses.

Similar to the spin-glass transition, the Gardner transition induces 
the split of basins on the free-energy landscape
and aging effects,  which suggests that the order parameter must be generalized: 
 one considers 
(i) the mean-squared displacement (MSD) $\Delta (\tau,t_{\rm w})$ and (ii) the distance between pairs of independently quenched configurations $\Delta_{AB} (t)$.
Here the MSD is defined as,
$
\Delta (\tau,t_{\rm w}) =
\frac{1}{N} \sum_{i=1}^{N} \langle|\vec{r}_i(\tau+ t_{\rm w}) - \vec{r}_i(t_{\rm w})|^2 \rangle
$, 
averaged over both thermal fluctuations and disorder, at the target $\varphi$ reached by compression. A waiting time $t_{\rm w}$ is introduced  in order to explicitly examine the aging effects (the total time $t$ is the sum of the measurement time $\tau$ and $t_{\rm w}$). 
On the other hand, 
$\Delta_{AB} (t) = \frac{1}{N}
\sum _{i=1}^{N} 
\langle|\vec{r}_i^A(t) - \vec{r}_i^B(t)|^2 \rangle
$,
where the two copies $A$ and $B$ are  independent  realizations at $\varphi$, compressed from the same initial sample at $\varphi_{\rm g}$.

The large-time limits of these quantities have important physical meanings.
The EA order parameter is defined as $\Delta_{\rm EA} \equiv \lim_{\tau \to \infty}\lim_{t_{\rm w} \to \infty} \Delta(\tau,t_{\rm w})$. 
Here the order of time limits is crucial \cite{bouchaud1998out}:
by reversing the order one can define another parameter,
$\Delta_{AB} \equiv \lim_{t_{\rm w} \to \infty}\lim_{\tau \to \infty}\Delta(\tau,t_{\rm w})=\lim_{t \to \infty}\Delta_{AB}(t)$.
The RSB  is signaled by $\Delta_{\rm AB} > \Delta_{\rm EA}$ (note that $\Delta_{AB}=\Delta_{\rm EA}$ in stable glasses).
In other words, the two large-time limits cannot be interchanged in the Gardner phase,
meaning that the aging effects become persistent.



In the second approach, one studies the response of hard-sphere glasses against a shear strain $\gamma$, analogous to observing magnetic susceptibilities in spin-glasses.
The simple strain $\gamma$ is applied to the $x$-coordinates of all particles ($x_i \to x_i + \gamma z_i$) after a waiting time $t_{\rm w}$, under the constant-volume and  Lees-Edwards boundary conditions~\cite{lees1972computer}.
The strain is increased  slowly with a  constant shear rate {\color{black} $\dot{\gamma}$},
and the reduced shear stress $\sigma=\beta \Sigma /\rho$
  is measured,
  where $\Sigma$ is the stress (for convenience, some data are presented with the  unitless stress rescaled by pressure, $\tilde{\sigma}=\sigma/p$).

As in the  spin-glass case,
one can consider two types of protocols, namely
{\it zero field compression} (ZFC) and {\it field compression} (FC)
\HY{(see Fig.~\ref{fig:landscape} (B))}.
In the  ZFC protocol, one compresses the configuration from $\varphi_{\rm g}$ to $\varphi$,  waits for time $t_{\rm w}$ before applying a shear  strain $\delta \gamma$ instantaneously, and then
measures the stress $\sigma_{\rm ZFC} (\tau, t_{\rm w})$ as a function of $\tau$. 
In the FC protocol, one applies
$\delta \gamma$ at the initial density $\varphi_{\rm g}$,  and 
then measures the stress $\sigma_{\rm FC}(t)$ once the configuration is compressed to $\varphi$ ($t$ is reset to zero after compression).
Similar to the caging order parameters, two large-time limits 
can be considered:
$\sigma_{\rm ZFC} \equiv  \lim_{\tau \to \infty}\lim_{t_{\rm w} \to \infty} \sigma_{\rm ZFC}(\tau,t_{\rm w})$ 
and $\sigma_{\rm FC} \equiv \lim_{t_{\rm w} \to \infty}\lim_{\tau \to \infty} \sigma_{\rm ZFC} (\tau, t_{\rm w})$.

Theories have demonstrated that the above two approaches (more specifically, the caging order parameters $\Delta$ and the shear moduli $\mu = \sigma/\delta \gamma$) are intrinsically related
~\cite{yoshino2014shear}: in the large pressure limit, 
 $\mu_{\rm ZFC} \sim 1/\Delta_{\rm EA}$ and $\mu_{\rm FC}/p \sim 1/\Delta_{AB}$. 
 These relationships are the counterpart of the duality between overlapping order parameters and magnetic  susceptibilities in spin-glasses.
  

\subsection{Aging effects}

\begin{figure}[h]
\centerline{\includegraphics[width=0.9\textwidth]{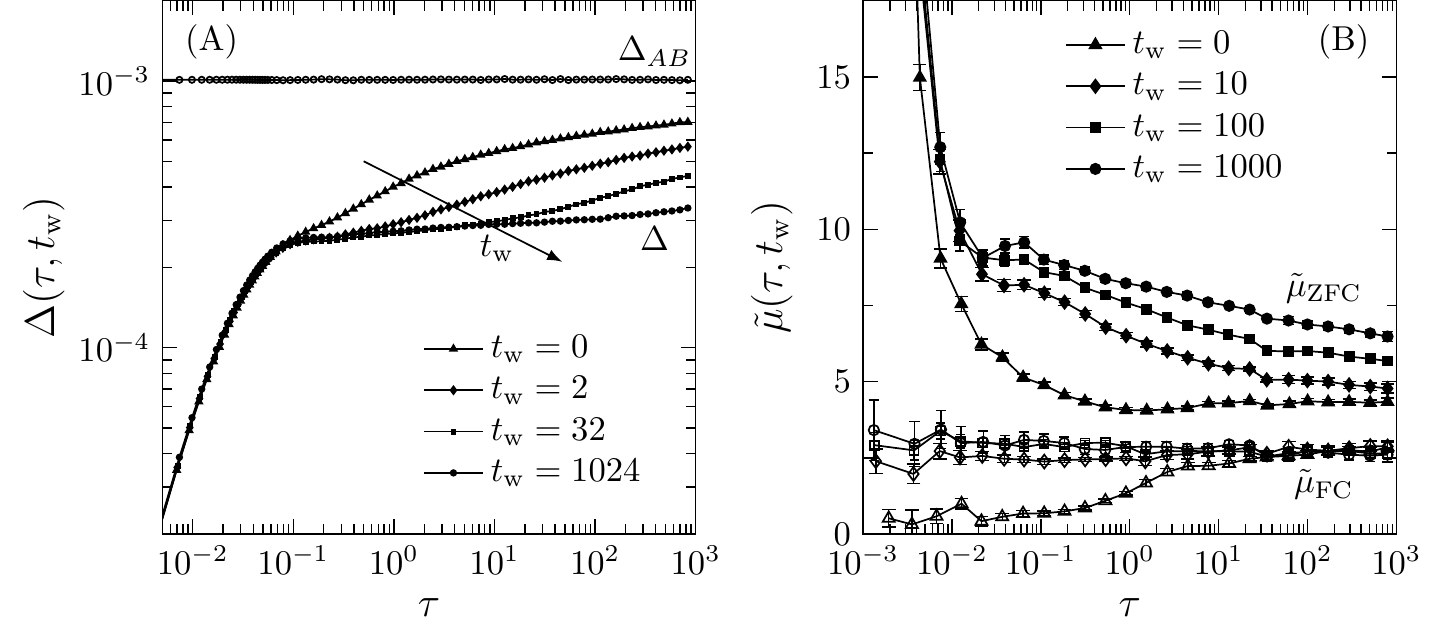}}
\caption{Time evolutions of (A) caging order parameters  
(adapted from~\cite{berthier2016growing})
and (B) shear moduli 
(adapted from~\cite{jin2017exploring}), in the Gardner phase of the hard-sphere glass model. 
}
\label{fig:aging}
\end{figure}

In the Gardner phase ($\varphi > \varphi_{\rm G}$), aging effects can be observed in both MSD (without shear deformations) and shear responses.
Figure~\ref{fig:aging}(A) shows the simulation data of MSD. 
After a short time  $\tau_{\rm b}\sim 1$ of ballistic motions, the evolution of $\Delta(\tau,t_{\rm w })$, as a function of $\tau$,
exhibits a plateau followed by further growth.
The switch from the former to the latter happens at longer times with   increasing  waiting time $t_{\rm w}$.
The height of the short-time plateau
gives  $\Delta_{\rm EA}$ (practically, we set $\Delta_{\rm EA} = \Delta (\tau=\tau_{\rm b},t_{\rm w}=0)$).
Figure~\ref{fig:aging}(A) also displays  $\Delta_{AB}(t )$, which is time-independent and should correspond to a long-time plateau of $\Delta(\tau,t_{\rm w })$ (this plateau is unfortunately beyond the current simulation time window). 
The clear separation of the two parameters ($\Delta_{AB} > \Delta_{\rm EA}$) is the first numerical evidence of the ergodicity breaking in the Gardner phase~\cite{charbonneau2015numerical, berthier2016growing,seoane2018spin}. 


Figure~\ref{fig:aging}(B) shows the time-dependent
(unitless) shear moduli, $\tilde{\mu}(\tau,t_{\rm w})$,
whose behavior is similar to that of
MSD.
An important feature 
is that $\tilde{\mu}_{\rm ZFC}(\tau,t_{\rm w})$ exhibits a plateau suggesting the existence of $\tilde{\mu}_{\rm ZFC}$.
On the other hand, $\tilde{\mu}_{\rm FC}(t)$ is essentially a constant in time $t$ (for $t > \tau_{\rm b}$), which  shall be denoted as 
$\tilde{\mu}_{\rm FC}$. 
In the proper order of  large-time limits, one expects that  $\tilde{\mu}_{\rm ZFC}(\tau,t_{\rm w})$ decays to $\tilde{\mu}_{\rm FC}$, as $\lim_{t_{\rm w} \to \infty}\lim_{\tau \to \infty} \tilde{\mu}_{\rm ZFC} (\tau, t_{\rm w})=\tilde{\mu}_{\rm FC}$, but the convergence becomes slower as  $t_{\rm w}$ increases. 
  Apparently $\tilde{\mu}_{\rm ZFC}$ is larger than $\tilde{\mu}_{\rm FC}$, 
  which parallels $\Delta_{AB} > \Delta_{\rm EA}$.

 
\subsection{Anomalous order parameters and responses}

\begin{figure}[h]
\centerline{\includegraphics[width=0.9\textwidth]{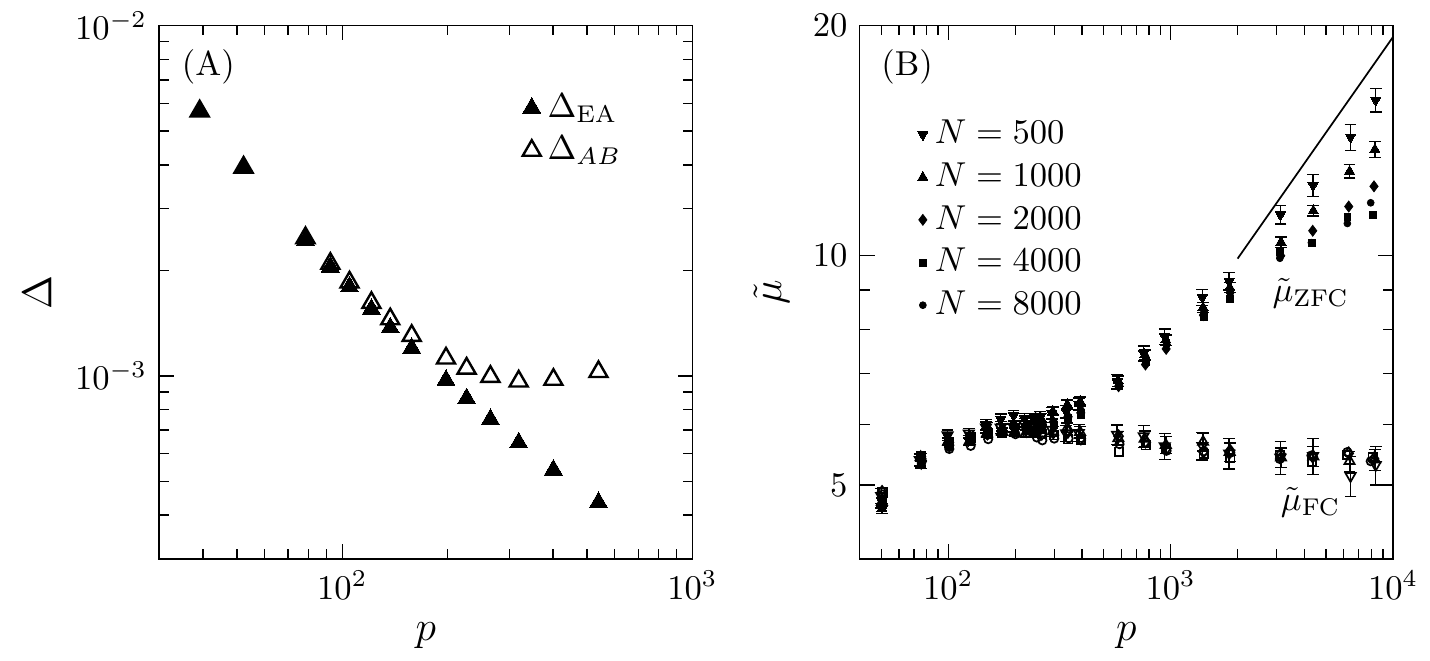}}
\caption{{\color{black} (A) Bifurcation of caging order parameters $\Delta_{\rm EA}$ and $\Delta_{AB}$ around the Gardner transition $p_{\rm G} \approx 2.7 \times 10^2$}
(adapted from~\cite{berthier2016growing}).
(B) Protocol-dependent shear moduli $\tilde{\mu}_{\rm ZFC}$ and $\tilde{\mu}_{\rm FC}$ (adapted from~\cite{jin2017exploring}).
The solid line indicates the  scaling
$\mu_{\rm ZFC} \sim p^{1.41574}$ predicted by the mean-field theory~\cite{yoshino2014shear}.
}
\label{fig:modulus}
\end{figure}

Figure~\ref{fig:modulus} shows the pressure dependence of caging order 
parameters ($\Delta_{\rm EA}$ and $\Delta_{AB}$) and shear moduli ($\tilde{\mu}_{\rm ZFC}$ and $\tilde{\mu}_{\rm FC}$) obtained through the above-mentioned dynamic 
measurements.
One finds that, in the stable glass phase ($p < p_{\rm G}$), $\Delta_{\rm EA}=\Delta_{\rm AB}$
and  $\tilde{\mu}_{\rm ZFC}=\tilde{\mu}_{\rm FC}$,
while in the Gardner phase ($p > p_{\rm G}$), $\Delta_{\rm EA}< \Delta_{\rm AB}$ and $\tilde{\mu}_{\rm ZFC} >\tilde{\mu}_{\rm FC}$.
In the large pressure limit, mean-field theories predict that $\Delta_{\rm EA} \sim p^{-\kappa}$~\cite{charbonneau2014fractal} and $\mu_{\rm ZFC} \sim p^\kappa$~\cite{yoshino2014shear}, where $\kappa = 1.41574$. The former is verified by three-dimensional simulations in Ref.~\cite{charbonneau2014fractal} and the latter by those in Ref.~\cite{jin2017exploring} (see Fig.~\ref{fig:modulus}(B)).
 The theories also give large-$p$ predictions $\mu_{\rm FC}/p \sim 1/\Delta_{AB} \sim \rm{constant}$, which are  consistent with the simulation results in Fig.~\ref{fig:modulus}.

\section{Gardner transition under shear}

\begin{figure}[h]
\centerline{\includegraphics[width=0.9\textwidth]{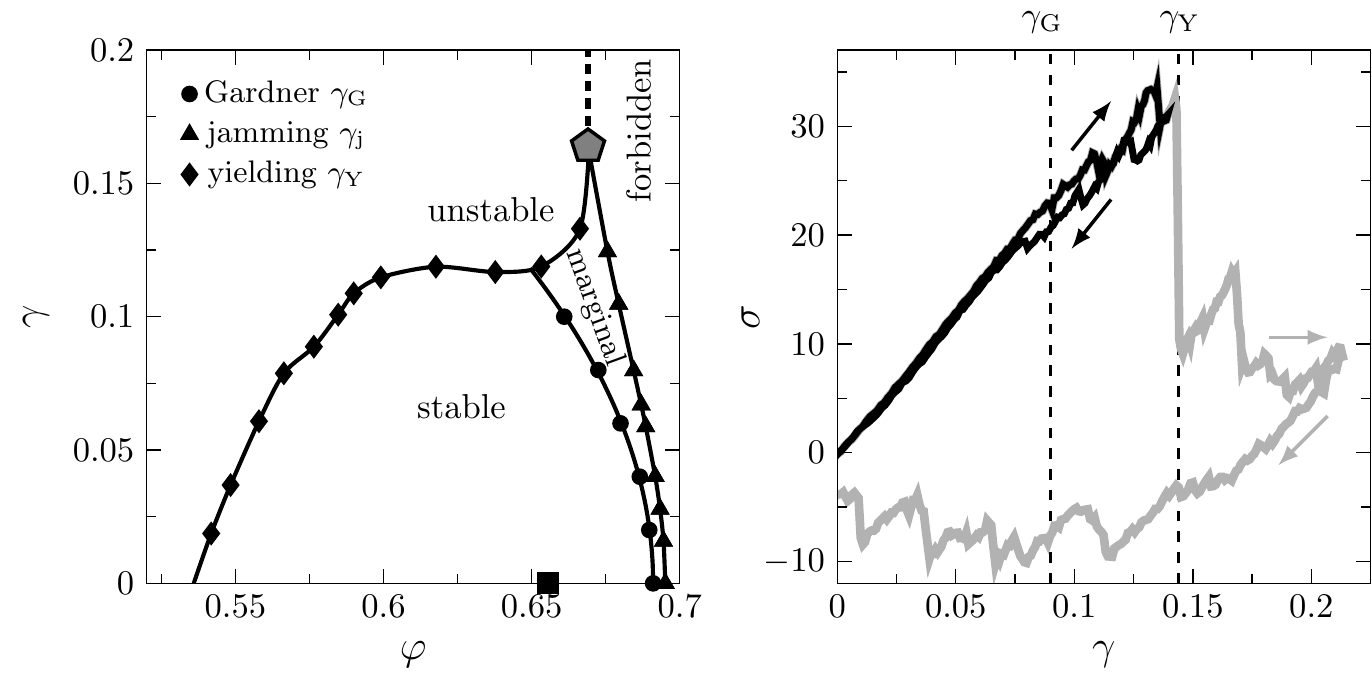}}
\caption{{\color{black} (A) Stability map of hard-sphere glasses, obtained from the initial glass at $\varphi = \varphi_{\rm g}$ and $\gamma = 0$ (square).}
 (B) Stress-strain curves 
 measured in cyclic shear simulations {\color{black} at a constant $\varphi=0.66$}.
Black and gray lines correspond to $\gamma_{\rm G} < \gamma_{\rm max} < \gamma_{\rm Y}$ and $\gamma_{\rm max} > \gamma_{\rm Y}$. {\color{black} Adapted from~\cite{jin2018stability}).}}
 \label{fig:cyclic_shear} 
\end{figure}

As predicted theoretically~{\cite{rainone2015following}}, 
a Gardner transition at $\gamma_{\rm G}$ could occur under shear,
before the glass yields at $\gamma_{\rm Y}$.
{\color{black} Figure~\ref{fig:cyclic_shear}(A) shows the {\it stability map} of hard sphere glasses under  shear and compression/decompression~\cite{jin2018stability, altieri2019mean}.}
The Gardner transition and yielding give rise to three types of behavior in a typical cyclic shear test (see {\color{black} Fig.~\ref{fig:cyclic_shear}(B)}). 
(i) The stress-strain curve is reversible in the stable glass phase ($\gamma < \gamma_{\rm G}$).
(ii) If the shear strain is reversed at a maximum strain $\gamma_{\rm max}$ between   $\gamma_{\rm G}$ and  $\gamma_{\rm Y}$, a hysteresis loop emerges, which however disappears below  $\gamma_{\rm G}$. This partial-reversible phenomenon  is a manifestation of the 
hierarchical free-energy landscape consisting of  basins within a common meta-basin.
The part of the stress-strain curve in the Gardner phase ($\gamma_{\rm G} < \gamma < \gamma_{\rm Y}$) is jerky
due to many small avalanches, reflecting the marginal stability. (iii) If $\gamma_{\rm max}> \gamma_{\rm Y}$, the cycle becomes strongly irreversible, suggesting the destruction of glass meta-basin after yielding.

\section{Discussion and outlook}

As a second-order phase transition, the fluctuation of order parameters (or the susceptibility) 
is expected to diverge at the Gardner transition in the thermodynamic limit. Simulations have shown that the caging susceptibility  grows orders of magnitude approaching the Gardner transition~\cite{berthier2016growing}.
Furthermore, the spatial correlations between local caging order parameters become long-ranged in the Gardner phase,  {\color{black} implying} the heterogeneity of vibrational dynamics~\cite{berthier2016growing, liao2019hierarchical}.
However, dynamical activations could possibly turn a mean-field thermodynamic phase transition into a crossover in low dimensions.
{\color{black} It remains inconclusive whether a sharp Gardner transition survives in three dimensions, although a renormalization group theory based on loop expansions~\cite{charbonneau2017nontrivial} (see Chapter 3 for details) and a machine-learning facilitated finite-size analysis of simulation data~\cite{li2021determining} seem to suggest so.}

The discussion so far has focused on the hard-sphere model.
Hard spheres 
have a well-defined singularity under compression, the jamming transition, where quantities such as  pressure and the length scale of mechanical response diverge {\color{black}(see Chapter 19 for a review on the jamming transition)}. Because the jamming transition lies in the Gardner phase, the full RSB predictions should also apply to 
the criticality and marginality of jamming, which are quantified by power-law scalings of weak forces, small interparticle gaps~\cite{charbonneau2014fractal} and low-frequency vibrational modes~\cite{franz2015universal}.
Remarkably, numerical results seemingly agree with the 
mean-field  exponents 
{ even in low dimensions~\cite{charbonneau2014fractal, charbonneau2015jamming, charbonneau2016universal}. The evidence of ultrametricity that characterizes the hierarchical energy landscape, has also been demonstrated numerically in jammed packings in three dimensions~\cite{dennis2020jamming}. }

{ The Gardner transition seems to emerge as a ``precursor"  of certain singularities (jamming under compression and yielding under shear) in hard particles. The situation is more complicated in cases without such singularities, e.g., cooling soft spheres under the constant density condition. On the one hand, the mean-field theory universally identifies the existence of Gardner transition in soft spheres~\cite{ scalliet2019marginally}, and simulations have reported a rejuvenation-memory effect~\cite{scalliet2019rejuvenation} similar to that found in spin-glasses.
On the other hand, however, simulations demonstrate that the Gardner transition could be interfered with by low-dimensional effects such as localized ``defects"~\cite{scalliet2017absence}. Separating the Gardner physics from strong low-dimensional effects in soft spheres remains a  challenge in simulations.}

Finally, experimental efforts to detect the Gardner transition have shown encouraging progress. The caging order parameter approach is applied to vibrated granular discs, providing evidence of the Gardner phase~\cite{seguin2016experimental}. In the Gardner phase, one expects a logarithmic growth of the MSD with lag time, which is verified in an experiment of glassy colloidal suspensions~\cite{hammond2020experimental}. The experimental data of shear modulus and  MSD in a hard-sphere colloidal glass are consistent with the scalings $\mu_{\rm ZFC} \sim 1/\Delta_{\rm EA} 
\sim p^{\kappa}$~\cite{zargar2017scaling}. The critical scalings of weak forces and small interparticle gaps have also been verified by precise experimental measurements of jammed photo-elastic disks~{\color{black}\cite{doi:10.1073/pnas.2204879119}}.
{\color{black} The evidence of a Gardner-like transition is reported in a two-dimensional bidisperse granular crystal~\cite{kool2022gardner}, suggesting that the Gardner physics could be observed with minimum disorder~\cite{charbonneau2019glassy}.} 
Examining protocol-dependent shear moduli and complex aging dynamics could provide future directions for the experimental exploration of Gardner physics.

\bibliography{Gardner_simulations}

\begin{thebibliography}{69}%
\makeatletter
\providecommand \@ifxundefined [1]{%
 \@ifx{#1\undefined}
}%
\providecommand \@ifnum [1]{%
 \ifnum #1\expandafter \@firstoftwo
 \else \expandafter \@secondoftwo
 \fi
}%
\providecommand \@ifx [1]{%
 \ifx #1\expandafter \@firstoftwo
 \else \expandafter \@secondoftwo
 \fi
}%
\providecommand \natexlab [1]{#1}%
\providecommand \enquote  [1]{``#1''}%
\providecommand \bibnamefont  [1]{#1}%
\providecommand \bibfnamefont [1]{#1}%
\providecommand \citenamefont [1]{#1}%
\providecommand \href@noop [0]{\@secondoftwo}%
\providecommand \href [0]{\begingroup \@sanitize@url \@href}%
\providecommand \@href[1]{\@@startlink{#1}\@@href}%
\providecommand \@@href[1]{\endgroup#1\@@endlink}%
\providecommand \@sanitize@url [0]{\catcode `\\12\catcode `\$12\catcode
  `\&12\catcode `\#12\catcode `\^12\catcode `\_12\catcode `\%12\relax}%
\providecommand \@@startlink[1]{}%
\providecommand \@@endlink[0]{}%
\providecommand \url  [0]{\begingroup\@sanitize@url \@url }%
\providecommand \@url [1]{\endgroup\@href {#1}{\urlprefix }}%
\providecommand \urlprefix  [0]{URL }%
\providecommand \Eprint [0]{\href }%
\providecommand \doibase [0]{http://dx.doi.org/}%
\providecommand \selectlanguage [0]{\@gobble}%
\providecommand \bibinfo  [0]{\@secondoftwo}%
\providecommand \bibfield  [0]{\@secondoftwo}%
\providecommand \translation [1]{[#1]}%
\providecommand \BibitemOpen [0]{}%
\providecommand \bibitemStop [0]{}%
\providecommand \bibitemNoStop [0]{.\EOS\space}%
\providecommand \EOS [0]{\spacefactor3000\relax}%
\providecommand \BibitemShut  [1]{\csname bibitem#1\endcsname}%
\let\auto@bib@innerbib\@empty
\bibitem [{\citenamefont {Parisi}\ \emph {et~al.}(2020)\citenamefont {Parisi},
  \citenamefont {Urbani},\ and\ \citenamefont {Zamponi}}]{parisi2020theory}%
  \BibitemOpen
  \bibfield  {author} {\bibinfo {author} {\bibfnamefont {Giorgio}\ \bibnamefont
  {Parisi}}, \bibinfo {author} {\bibfnamefont {Pierfrancesco}\ \bibnamefont
  {Urbani}}, \ and\ \bibinfo {author} {\bibfnamefont {Francesco}\ \bibnamefont
  {Zamponi}},\ }\href@noop {} {\emph {\bibinfo {title} {Theory of Simple
  Glasses: Exact Solutions in Infinite Dimensions}}}\ (\bibinfo  {publisher}
  {Cambridge University Press},\ \bibinfo {year} {2020})\BibitemShut {NoStop}%
\bibitem [{\citenamefont {Berthier}\ \emph {et~al.}(2019)\citenamefont
  {Berthier}, \citenamefont {Biroli}, \citenamefont {Charbonneau},
  \citenamefont {Corwin}, \citenamefont {Franz},\ and\ \citenamefont
  {Zamponi}}]{berthier2019gardner}%
  \BibitemOpen
  \bibfield  {author} {\bibinfo {author} {\bibfnamefont {Ludovic}\ \bibnamefont
  {Berthier}}, \bibinfo {author} {\bibfnamefont {Giulio}\ \bibnamefont
  {Biroli}}, \bibinfo {author} {\bibfnamefont {Patrick}\ \bibnamefont
  {Charbonneau}}, \bibinfo {author} {\bibfnamefont {Eric~I}\ \bibnamefont
  {Corwin}}, \bibinfo {author} {\bibfnamefont {Silvio}\ \bibnamefont {Franz}},
  \ and\ \bibinfo {author} {\bibfnamefont {Francesco}\ \bibnamefont
  {Zamponi}},\ }\bibfield  {title} {\enquote {\bibinfo {title} {Gardner physics
  in amorphous solids and beyond},}\ }\href {\doibase 10.1063/1.5097175}
  {\bibfield  {journal} {\bibinfo  {journal} {The Journal of chemical physics}\
  }\textbf {\bibinfo {volume} {151}},\ \bibinfo {pages} {010901} (\bibinfo
  {year} {2019})}\BibitemShut {NoStop}%
\bibitem [{\citenamefont {M\'ezard}\ \emph {et~al.}(1987)\citenamefont
  {M\'ezard}, \citenamefont {Parisi},\ and\ \citenamefont {Virasoro}}]{MPV87}%
  \BibitemOpen
  \bibfield  {author} {\bibinfo {author} {\bibfnamefont {M.}~\bibnamefont
  {M\'ezard}}, \bibinfo {author} {\bibfnamefont {G.}~\bibnamefont {Parisi}}, \
  and\ \bibinfo {author} {\bibfnamefont {M.~A.}\ \bibnamefont {Virasoro}},\
  }\href@noop {} {\emph {\bibinfo {title} {Spin glass theory and beyond}}}\
  (\bibinfo  {publisher} {World Scientific},\ \bibinfo {address} {Singapore},\
  \bibinfo {year} {1987})\BibitemShut {NoStop}%
\bibitem [{\citenamefont {Nagata}\ \emph {et~al.}(1979)\citenamefont {Nagata},
  \citenamefont {Keesom},\ and\ \citenamefont {Harrison}}]{nagata1979low}%
  \BibitemOpen
  \bibfield  {author} {\bibinfo {author} {\bibfnamefont {Shoichi}\ \bibnamefont
  {Nagata}}, \bibinfo {author} {\bibfnamefont {PH}~\bibnamefont {Keesom}}, \
  and\ \bibinfo {author} {\bibfnamefont {HR}~\bibnamefont {Harrison}},\
  }\bibfield  {title} {\enquote {\bibinfo {title} {Low-dc-field susceptibility
  of cu mn spin glass},}\ }\href@noop {} {\bibfield  {journal} {\bibinfo
  {journal} {Physical Review B}\ }\textbf {\bibinfo {volume} {19}},\ \bibinfo
  {pages} {1633} (\bibinfo {year} {1979})}\BibitemShut {NoStop}%
\bibitem [{\citenamefont {Edwards}\ and\ \citenamefont
  {Anderson}(1975)}]{edwards1975theory}%
  \BibitemOpen
  \bibfield  {author} {\bibinfo {author} {\bibfnamefont {Samuel~Frederick}\
  \bibnamefont {Edwards}}\ and\ \bibinfo {author} {\bibfnamefont {Phil~W}\
  \bibnamefont {Anderson}},\ }\bibfield  {title} {\enquote {\bibinfo {title}
  {Theory of spin glasses},}\ }\href@noop {} {\bibfield  {journal} {\bibinfo
  {journal} {Journal of Physics F: Metal Physics}\ }\textbf {\bibinfo {volume}
  {5}},\ \bibinfo {pages} {965} (\bibinfo {year} {1975})}\BibitemShut {NoStop}%
\bibitem [{\citenamefont {Castellani}\ and\ \citenamefont
  {Cavagna}(2005)}]{castellani2005spin}%
  \BibitemOpen
  \bibfield  {author} {\bibinfo {author} {\bibfnamefont {Tommaso}\ \bibnamefont
  {Castellani}}\ and\ \bibinfo {author} {\bibfnamefont {Andrea}\ \bibnamefont
  {Cavagna}},\ }\bibfield  {title} {\enquote {\bibinfo {title} {Spin-glass
  theory for pedestrians},}\ }\href@noop {} {\bibfield  {journal} {\bibinfo
  {journal} {Journal of Statistical Mechanics: Theory and Experiment}\ }\textbf
  {\bibinfo {volume} {2005}},\ \bibinfo {pages} {P05012} (\bibinfo {year}
  {2005})}\BibitemShut {NoStop}%
\bibitem [{\citenamefont {Granberg}\ \emph {et~al.}(1988)\citenamefont
  {Granberg}, \citenamefont {Sandlund}, \citenamefont {Nordblad}, \citenamefont
  {Svedlindh},\ and\ \citenamefont {Lundgren}}]{granberg1988observation}%
  \BibitemOpen
  \bibfield  {author} {\bibinfo {author} {\bibfnamefont {P}~\bibnamefont
  {Granberg}}, \bibinfo {author} {\bibfnamefont {L}~\bibnamefont {Sandlund}},
  \bibinfo {author} {\bibfnamefont {P}~\bibnamefont {Nordblad}}, \bibinfo
  {author} {\bibfnamefont {P}~\bibnamefont {Svedlindh}}, \ and\ \bibinfo
  {author} {\bibfnamefont {L}~\bibnamefont {Lundgren}},\ }\bibfield  {title}
  {\enquote {\bibinfo {title} {Observation of a time-dependent spatial
  correlation length in a metallic spin glass},}\ }\href@noop {} {\bibfield
  {journal} {\bibinfo  {journal} {Physical Review B}\ }\textbf {\bibinfo
  {volume} {38}},\ \bibinfo {pages} {7097} (\bibinfo {year}
  {1988})}\BibitemShut {NoStop}%
\bibitem [{\citenamefont {Vincent}\ \emph {et~al.}(1997)\citenamefont
  {Vincent}, \citenamefont {Hammann}, \citenamefont {Ocio}, \citenamefont
  {Bouchaud},\ and\ \citenamefont {Cugliandolo}}]{vincent1997slow}%
  \BibitemOpen
  \bibfield  {author} {\bibinfo {author} {\bibfnamefont {Eric}\ \bibnamefont
  {Vincent}}, \bibinfo {author} {\bibfnamefont {Jacques}\ \bibnamefont
  {Hammann}}, \bibinfo {author} {\bibfnamefont {Miguel}\ \bibnamefont {Ocio}},
  \bibinfo {author} {\bibfnamefont {Jean-Philippe}\ \bibnamefont {Bouchaud}}, \
  and\ \bibinfo {author} {\bibfnamefont {Leticia~F}\ \bibnamefont
  {Cugliandolo}},\ }\bibfield  {title} {\enquote {\bibinfo {title} {Slow
  dynamics and aging in spin glasses},}\ }in\ \href@noop {} {\emph {\bibinfo
  {booktitle} {Complex Behaviour of Glassy Systems}}}\ (\bibinfo  {publisher}
  {Springer},\ \bibinfo {year} {1997})\ pp.\ \bibinfo {pages}
  {184--219}\BibitemShut {NoStop}%
\bibitem [{\citenamefont {Nordblad}\ and\ \citenamefont
  {Svedlindh}(1998)}]{nordblad1998experiments}%
  \BibitemOpen
  \bibfield  {author} {\bibinfo {author} {\bibfnamefont {Per}\ \bibnamefont
  {Nordblad}}\ and\ \bibinfo {author} {\bibfnamefont {Peter}\ \bibnamefont
  {Svedlindh}},\ }\bibfield  {title} {\enquote {\bibinfo {title} {Experiments
  on spin glasses},}\ }in\ \href@noop {} {\emph {\bibinfo {booktitle} {Spin
  Glasses and Random Fields}}}\ (\bibinfo  {publisher} {World Scientific},\
  \bibinfo {year} {1998})\ pp.\ \bibinfo {pages} {1--27}\BibitemShut {NoStop}%
\bibitem [{\citenamefont {Cugliandolo}\ and\ \citenamefont
  {Kurchan}(1993)}]{cugliandolo1993analytical}%
  \BibitemOpen
  \bibfield  {author} {\bibinfo {author} {\bibfnamefont {Leticia~F}\
  \bibnamefont {Cugliandolo}}\ and\ \bibinfo {author} {\bibfnamefont {Jorge}\
  \bibnamefont {Kurchan}},\ }\bibfield  {title} {\enquote {\bibinfo {title}
  {Analytical solution of the off-equilibrium dynamics of a long-range
  spin-glass model},}\ }\href {\doibase 10.1103/PhysRevLett.71.173} {\bibfield
  {journal} {\bibinfo  {journal} {Physical Review Letters}\ }\textbf {\bibinfo
  {volume} {71}},\ \bibinfo {pages} {173} (\bibinfo {year} {1993})}\BibitemShut
  {NoStop}%
\bibitem [{\citenamefont {Cugliandolo}\ and\ \citenamefont
  {Kurchan}(1994)}]{cugliandolo1994out}%
  \BibitemOpen
  \bibfield  {author} {\bibinfo {author} {\bibfnamefont {Leticia~F}\
  \bibnamefont {Cugliandolo}}\ and\ \bibinfo {author} {\bibfnamefont {Jorge}\
  \bibnamefont {Kurchan}},\ }\bibfield  {title} {\enquote {\bibinfo {title} {On
  the out-of-equilibrium relaxation of the sherrington-kirkpatrick model},}\
  }\href {\doibase 10.1088/0305-4470/27/17/011} {\bibfield  {journal} {\bibinfo
   {journal} {Journal of Physics A: Mathematical and General}\ }\textbf
  {\bibinfo {volume} {27}},\ \bibinfo {pages} {5749} (\bibinfo {year}
  {1994})}\BibitemShut {NoStop}%
\bibitem [{\citenamefont {Cugliandolo}\ \emph {et~al.}(1997)\citenamefont
  {Cugliandolo}, \citenamefont {Kurchan},\ and\ \citenamefont
  {Peliti}}]{cugliandolo1997energy}%
  \BibitemOpen
  \bibfield  {author} {\bibinfo {author} {\bibfnamefont {Leticia~F}\
  \bibnamefont {Cugliandolo}}, \bibinfo {author} {\bibfnamefont {Jorge}\
  \bibnamefont {Kurchan}}, \ and\ \bibinfo {author} {\bibfnamefont {Luca}\
  \bibnamefont {Peliti}},\ }\bibfield  {title} {\enquote {\bibinfo {title}
  {Energy flow, partial equilibration, and effective temperatures in systems
  with slow dynamics},}\ }\href@noop {} {\bibfield  {journal} {\bibinfo
  {journal} {Physical Review E}\ }\textbf {\bibinfo {volume} {55}},\ \bibinfo
  {pages} {3898} (\bibinfo {year} {1997})}\BibitemShut {NoStop}%
\bibitem [{\citenamefont {Franz}\ \emph {et~al.}(1998)\citenamefont {Franz},
  \citenamefont {M{\'e}zard}, \citenamefont {Parisi},\ and\ \citenamefont
  {Peliti}}]{franz1998measuring}%
  \BibitemOpen
  \bibfield  {author} {\bibinfo {author} {\bibfnamefont {Silvio}\ \bibnamefont
  {Franz}}, \bibinfo {author} {\bibfnamefont {Marc}\ \bibnamefont
  {M{\'e}zard}}, \bibinfo {author} {\bibfnamefont {Giorgio}\ \bibnamefont
  {Parisi}}, \ and\ \bibinfo {author} {\bibfnamefont {Luca}\ \bibnamefont
  {Peliti}},\ }\bibfield  {title} {\enquote {\bibinfo {title} {Measuring
  equilibrium properties in aging systems},}\ }\href@noop {} {\bibfield
  {journal} {\bibinfo  {journal} {Physical Review Letters}\ }\textbf {\bibinfo
  {volume} {81}},\ \bibinfo {pages} {1758} (\bibinfo {year}
  {1998})}\BibitemShut {NoStop}%
\bibitem [{\citenamefont {Marinari}\ \emph {et~al.}(2000)\citenamefont
  {Marinari}, \citenamefont {Parisi}, \citenamefont {Ricci-Tersenghi},\ and\
  \citenamefont {Ruiz-Lorenzo}}]{marinari2000off}%
  \BibitemOpen
  \bibfield  {author} {\bibinfo {author} {\bibfnamefont {Enzo}\ \bibnamefont
  {Marinari}}, \bibinfo {author} {\bibfnamefont {Giorgio}\ \bibnamefont
  {Parisi}}, \bibinfo {author} {\bibfnamefont {Federico}\ \bibnamefont
  {Ricci-Tersenghi}}, \ and\ \bibinfo {author} {\bibfnamefont {Juan~J}\
  \bibnamefont {Ruiz-Lorenzo}},\ }\bibfield  {title} {\enquote {\bibinfo
  {title} {Off-equilibrium dynamics at very low temperatures in
  three-dimensional spin glasses},}\ }\href@noop {} {\bibfield  {journal}
  {\bibinfo  {journal} {Journal of Physics A: Mathematical and General}\
  }\textbf {\bibinfo {volume} {33}},\ \bibinfo {pages} {2373} (\bibinfo {year}
  {2000})}\BibitemShut {NoStop}%
\bibitem [{\citenamefont {Yoshino}\ \emph {et~al.}(2002)\citenamefont
  {Yoshino}, \citenamefont {Hukushima},\ and\ \citenamefont
  {Takayama}}]{yoshino2002extended}%
  \BibitemOpen
  \bibfield  {author} {\bibinfo {author} {\bibfnamefont {Hajime}\ \bibnamefont
  {Yoshino}}, \bibinfo {author} {\bibfnamefont {Koji}\ \bibnamefont
  {Hukushima}}, \ and\ \bibinfo {author} {\bibfnamefont {Hajime}\ \bibnamefont
  {Takayama}},\ }\bibfield  {title} {\enquote {\bibinfo {title} {Extended
  droplet theory for aging in short-range spin glasses and a numerical
  examination},}\ }\href@noop {} {\bibfield  {journal} {\bibinfo  {journal}
  {Physical Review B}\ }\textbf {\bibinfo {volume} {66}},\ \bibinfo {pages}
  {064431} (\bibinfo {year} {2002})}\BibitemShut {NoStop}%
\bibitem [{\citenamefont {Bray}\ and\ \citenamefont
  {Moore}(1987)}]{bray1987chaotic}%
  \BibitemOpen
  \bibfield  {author} {\bibinfo {author} {\bibfnamefont {Alan~J}\ \bibnamefont
  {Bray}}\ and\ \bibinfo {author} {\bibfnamefont {Michael~A}\ \bibnamefont
  {Moore}},\ }\bibfield  {title} {\enquote {\bibinfo {title} {Chaotic nature of
  the spin-glass phase},}\ }\href@noop {} {\bibfield  {journal} {\bibinfo
  {journal} {Physical review letters}\ }\textbf {\bibinfo {volume} {58}},\
  \bibinfo {pages} {57} (\bibinfo {year} {1987})}\BibitemShut {NoStop}%
\bibitem [{\citenamefont {Fisher}\ and\ \citenamefont
  {Huse}(1988{\natexlab{a}})}]{fisher1988equilibrium}%
  \BibitemOpen
  \bibfield  {author} {\bibinfo {author} {\bibfnamefont {Daniel~S}\
  \bibnamefont {Fisher}}\ and\ \bibinfo {author} {\bibfnamefont {David~A}\
  \bibnamefont {Huse}},\ }\bibfield  {title} {\enquote {\bibinfo {title}
  {Equilibrium behavior of the spin-glass ordered phase},}\ }\href@noop {}
  {\bibfield  {journal} {\bibinfo  {journal} {Physical Review B}\ }\textbf
  {\bibinfo {volume} {38}},\ \bibinfo {pages} {386} (\bibinfo {year}
  {1988}{\natexlab{a}})}\BibitemShut {NoStop}%
\bibitem [{\citenamefont {Fisher}\ and\ \citenamefont
  {Huse}(1988{\natexlab{b}})}]{fisher1988nonequilibrium}%
  \BibitemOpen
  \bibfield  {author} {\bibinfo {author} {\bibfnamefont {Daniel~S}\
  \bibnamefont {Fisher}}\ and\ \bibinfo {author} {\bibfnamefont {David~A}\
  \bibnamefont {Huse}},\ }\bibfield  {title} {\enquote {\bibinfo {title}
  {Nonequilibrium dynamics of spin glasses},}\ }\href@noop {} {\bibfield
  {journal} {\bibinfo  {journal} {Physical Review B}\ }\textbf {\bibinfo
  {volume} {38}},\ \bibinfo {pages} {373} (\bibinfo {year}
  {1988}{\natexlab{b}})}\BibitemShut {NoStop}%
\bibitem [{\citenamefont {Kondor}(1989)}]{kondor1989chaos}%
  \BibitemOpen
  \bibfield  {author} {\bibinfo {author} {\bibfnamefont {I}~\bibnamefont
  {Kondor}},\ }\bibfield  {title} {\enquote {\bibinfo {title} {On chaos in spin
  glasses},}\ }\href@noop {} {\bibfield  {journal} {\bibinfo  {journal}
  {Journal of Physics A: Mathematical and General}\ }\textbf {\bibinfo {volume}
  {22}},\ \bibinfo {pages} {L163} (\bibinfo {year} {1989})}\BibitemShut
  {NoStop}%
\bibitem [{\citenamefont {Rizzo}\ and\ \citenamefont
  {Crisanti}(2003)}]{rizzo2003chaos}%
  \BibitemOpen
  \bibfield  {author} {\bibinfo {author} {\bibfnamefont {Tommaso}\ \bibnamefont
  {Rizzo}}\ and\ \bibinfo {author} {\bibfnamefont {Andrea}\ \bibnamefont
  {Crisanti}},\ }\bibfield  {title} {\enquote {\bibinfo {title} {Chaos in
  temperature in the sherrington-kirkpatrick model},}\ }\href@noop {}
  {\bibfield  {journal} {\bibinfo  {journal} {Physical review letters}\
  }\textbf {\bibinfo {volume} {90}},\ \bibinfo {pages} {137201} (\bibinfo
  {year} {2003})}\BibitemShut {NoStop}%
\bibitem [{\citenamefont {Rizzo}\ and\ \citenamefont
  {Yoshino}(2006)}]{rizzo2006chaos}%
  \BibitemOpen
  \bibfield  {author} {\bibinfo {author} {\bibfnamefont {Tommaso}\ \bibnamefont
  {Rizzo}}\ and\ \bibinfo {author} {\bibfnamefont {Hajime}\ \bibnamefont
  {Yoshino}},\ }\bibfield  {title} {\enquote {\bibinfo {title} {Chaos in glassy
  systems from a thouless-anderson-palmer perspective},}\ }\href@noop {}
  {\bibfield  {journal} {\bibinfo  {journal} {Physical Review B}\ }\textbf
  {\bibinfo {volume} {73}},\ \bibinfo {pages} {064416} (\bibinfo {year}
  {2006})}\BibitemShut {NoStop}%
\bibitem [{\citenamefont {Yoshino}\ and\ \citenamefont
  {Rizzo}(2008)}]{yoshino2008stepwise}%
  \BibitemOpen
  \bibfield  {author} {\bibinfo {author} {\bibfnamefont {Hajime}\ \bibnamefont
  {Yoshino}}\ and\ \bibinfo {author} {\bibfnamefont {Tommaso}\ \bibnamefont
  {Rizzo}},\ }\bibfield  {title} {\enquote {\bibinfo {title} {Stepwise
  responses in mesoscopic glassy systems: A mean-field approach},}\ }\href@noop
  {} {\bibfield  {journal} {\bibinfo  {journal} {Physical Review B}\ }\textbf
  {\bibinfo {volume} {77}},\ \bibinfo {pages} {104429} (\bibinfo {year}
  {2008})}\BibitemShut {NoStop}%
\bibitem [{\citenamefont {Parisi}\ and\ \citenamefont
  {Rizzo}(2010)}]{parisi2010chaos}%
  \BibitemOpen
  \bibfield  {author} {\bibinfo {author} {\bibfnamefont {Giorgio}\ \bibnamefont
  {Parisi}}\ and\ \bibinfo {author} {\bibfnamefont {Tommaso}\ \bibnamefont
  {Rizzo}},\ }\bibfield  {title} {\enquote {\bibinfo {title} {Chaos in
  temperature in diluted mean-field spin-glass},}\ }\href@noop {} {\bibfield
  {journal} {\bibinfo  {journal} {Journal of Physics A: Mathematical and
  Theoretical}\ }\textbf {\bibinfo {volume} {43}},\ \bibinfo {pages} {235003}
  (\bibinfo {year} {2010})}\BibitemShut {NoStop}%
\bibitem [{\citenamefont {Le~Doussal}\ \emph {et~al.}(2010)\citenamefont
  {Le~Doussal}, \citenamefont {M{\"u}ller},\ and\ \citenamefont
  {Wiese}}]{le2010avalanches}%
  \BibitemOpen
  \bibfield  {author} {\bibinfo {author} {\bibfnamefont {Pierre}\ \bibnamefont
  {Le~Doussal}}, \bibinfo {author} {\bibfnamefont {Markus}\ \bibnamefont
  {M{\"u}ller}}, \ and\ \bibinfo {author} {\bibfnamefont {Kay~J{\"o}rg}\
  \bibnamefont {Wiese}},\ }\bibfield  {title} {\enquote {\bibinfo {title}
  {Avalanches in mean-field models and the barkhausen noise in spin-glasses},}\
  }\href@noop {} {\bibfield  {journal} {\bibinfo  {journal} {EPL (Europhysics
  Letters)}\ }\textbf {\bibinfo {volume} {91}},\ \bibinfo {pages} {57004}
  (\bibinfo {year} {2010})}\BibitemShut {NoStop}%
\bibitem [{\citenamefont {Le~Doussal}\ \emph {et~al.}(2012)\citenamefont
  {Le~Doussal}, \citenamefont {M{\"u}ller},\ and\ \citenamefont
  {Wiese}}]{le2012equilibrium}%
  \BibitemOpen
  \bibfield  {author} {\bibinfo {author} {\bibfnamefont {Pierre}\ \bibnamefont
  {Le~Doussal}}, \bibinfo {author} {\bibfnamefont {Markus}\ \bibnamefont
  {M{\"u}ller}}, \ and\ \bibinfo {author} {\bibfnamefont {Kay~J{\"o}rg}\
  \bibnamefont {Wiese}},\ }\bibfield  {title} {\enquote {\bibinfo {title}
  {Equilibrium avalanches in spin glasses},}\ }\href@noop {} {\bibfield
  {journal} {\bibinfo  {journal} {Physical Review B}\ }\textbf {\bibinfo
  {volume} {85}},\ \bibinfo {pages} {214402} (\bibinfo {year}
  {2012})}\BibitemShut {NoStop}%
\bibitem [{\citenamefont {Franz}\ and\ \citenamefont
  {Spigler}(2017)}]{franz2017mean}%
  \BibitemOpen
  \bibfield  {author} {\bibinfo {author} {\bibfnamefont {Silvio}\ \bibnamefont
  {Franz}}\ and\ \bibinfo {author} {\bibfnamefont {Stefano}\ \bibnamefont
  {Spigler}},\ }\bibfield  {title} {\enquote {\bibinfo {title} {Mean-field
  avalanches in jammed spheres},}\ }\href@noop {} {\bibfield  {journal}
  {\bibinfo  {journal} {Physical Review E}\ }\textbf {\bibinfo {volume} {95}},\
  \bibinfo {pages} {022139} (\bibinfo {year} {2017})}\BibitemShut {NoStop}%
\bibitem [{\citenamefont {Jonason}\ \emph {et~al.}(1998)\citenamefont
  {Jonason}, \citenamefont {Vincent}, \citenamefont {Hammann}, \citenamefont
  {Bouchaud},\ and\ \citenamefont {Nordblad}}]{jonason1998memory}%
  \BibitemOpen
  \bibfield  {author} {\bibinfo {author} {\bibfnamefont {K}~\bibnamefont
  {Jonason}}, \bibinfo {author} {\bibfnamefont {E}~\bibnamefont {Vincent}},
  \bibinfo {author} {\bibfnamefont {J}~\bibnamefont {Hammann}}, \bibinfo
  {author} {\bibfnamefont {JP}~\bibnamefont {Bouchaud}}, \ and\ \bibinfo
  {author} {\bibfnamefont {P}~\bibnamefont {Nordblad}},\ }\bibfield  {title}
  {\enquote {\bibinfo {title} {Memory and chaos effects in spin glasses},}\
  }\href@noop {} {\bibfield  {journal} {\bibinfo  {journal} {Physical Review
  Letters}\ }\textbf {\bibinfo {volume} {81}},\ \bibinfo {pages} {3243}
  (\bibinfo {year} {1998})}\BibitemShut {NoStop}%
\bibitem [{\citenamefont {Yoshino}\ \emph {et~al.}(2001)\citenamefont
  {Yoshino}, \citenamefont {Lema{\i}tre},\ and\ \citenamefont
  {Bouchaud}}]{yoshino2001multiple}%
  \BibitemOpen
  \bibfield  {author} {\bibinfo {author} {\bibfnamefont {Hajime}\ \bibnamefont
  {Yoshino}}, \bibinfo {author} {\bibfnamefont {Ana{\"e}l}\ \bibnamefont
  {Lema{\i}tre}}, \ and\ \bibinfo {author} {\bibfnamefont {J-P}\ \bibnamefont
  {Bouchaud}},\ }\bibfield  {title} {\enquote {\bibinfo {title} {Multiple
  domain growth and memory in the droplet model for spin-glasses},}\
  }\href@noop {} {\bibfield  {journal} {\bibinfo  {journal} {The European
  Physical Journal B-Condensed Matter and Complex Systems}\ }\textbf {\bibinfo
  {volume} {20}},\ \bibinfo {pages} {367--395} (\bibinfo {year}
  {2001})}\BibitemShut {NoStop}%
\bibitem [{\citenamefont {J{\"o}nsson}\ \emph {et~al.}(2004)\citenamefont
  {J{\"o}nsson}, \citenamefont {Mathieu}, \citenamefont {Nordblad},
  \citenamefont {Yoshino}, \citenamefont {Katori},\ and\ \citenamefont
  {Ito}}]{jonsson2004nonequilibrium}%
  \BibitemOpen
  \bibfield  {author} {\bibinfo {author} {\bibfnamefont {PE}~\bibnamefont
  {J{\"o}nsson}}, \bibinfo {author} {\bibfnamefont {R}~\bibnamefont {Mathieu}},
  \bibinfo {author} {\bibfnamefont {Per}\ \bibnamefont {Nordblad}}, \bibinfo
  {author} {\bibfnamefont {H}~\bibnamefont {Yoshino}}, \bibinfo {author}
  {\bibfnamefont {H~Aruga}\ \bibnamefont {Katori}}, \ and\ \bibinfo {author}
  {\bibfnamefont {A}~\bibnamefont {Ito}},\ }\bibfield  {title} {\enquote
  {\bibinfo {title} {Nonequilibrium dynamics of spin glasses: Examination of
  the ghost domain scenario},}\ }\href@noop {} {\bibfield  {journal} {\bibinfo
  {journal} {Physical Review B}\ }\textbf {\bibinfo {volume} {70}},\ \bibinfo
  {pages} {174402} (\bibinfo {year} {2004})}\BibitemShut {NoStop}%
\bibitem [{\citenamefont {Yoshino}\ and\ \citenamefont
  {M{\'e}zard}(2010)}]{yoshino2010emergence}%
  \BibitemOpen
  \bibfield  {author} {\bibinfo {author} {\bibfnamefont {Hajime}\ \bibnamefont
  {Yoshino}}\ and\ \bibinfo {author} {\bibfnamefont {Marc}\ \bibnamefont
  {M{\'e}zard}},\ }\bibfield  {title} {\enquote {\bibinfo {title} {Emergence of
  rigidity at the structural glass transition: A first-principles
  computation},}\ }\href@noop {} {\bibfield  {journal} {\bibinfo  {journal}
  {Physical review letters}\ }\textbf {\bibinfo {volume} {105}},\ \bibinfo
  {pages} {015504} (\bibinfo {year} {2010})}\BibitemShut {NoStop}%
\bibitem [{\citenamefont {Yoshino}(2012)}]{YO12}%
  \BibitemOpen
  \bibfield  {author} {\bibinfo {author} {\bibfnamefont {Hajime}\ \bibnamefont
  {Yoshino}},\ }\bibfield  {title} {\enquote {\bibinfo {title} {Replica theory
  of the rigidity of structural glasses},}\ }\href {\doibase
  doi.org/10.1063/1.4722343} {\bibfield  {journal} {\bibinfo  {journal} {The
  Journal of Chemical Physics}\ }\textbf {\bibinfo {volume} {136}},\ \bibinfo
  {pages} {214108} (\bibinfo {year} {2012})}\BibitemShut {NoStop}%
\bibitem [{\citenamefont {Yoshino}\ and\ \citenamefont
  {Zamponi}(2014)}]{yoshino2014shear}%
  \BibitemOpen
  \bibfield  {author} {\bibinfo {author} {\bibfnamefont {Hajime}\ \bibnamefont
  {Yoshino}}\ and\ \bibinfo {author} {\bibfnamefont {Francesco}\ \bibnamefont
  {Zamponi}},\ }\bibfield  {title} {\enquote {\bibinfo {title} {Shear modulus
  of glasses: Results from the full replica-symmetry-breaking solution},}\
  }\href@noop {} {\bibfield  {journal} {\bibinfo  {journal} {Physical Review
  E}\ }\textbf {\bibinfo {volume} {90}},\ \bibinfo {pages} {022302} (\bibinfo
  {year} {2014})}\BibitemShut {NoStop}%
\bibitem [{\citenamefont {Maimbourg}\ \emph {et~al.}(2016)\citenamefont
  {Maimbourg}, \citenamefont {Kurchan},\ and\ \citenamefont
  {Zamponi}}]{maimbourg2016solution}%
  \BibitemOpen
  \bibfield  {author} {\bibinfo {author} {\bibfnamefont {Thibaud}\ \bibnamefont
  {Maimbourg}}, \bibinfo {author} {\bibfnamefont {Jorge}\ \bibnamefont
  {Kurchan}}, \ and\ \bibinfo {author} {\bibfnamefont {Francesco}\ \bibnamefont
  {Zamponi}},\ }\bibfield  {title} {\enquote {\bibinfo {title} {Solution of the
  dynamics of liquids in the large-dimensional limit},}\ }\href@noop {}
  {\bibfield  {journal} {\bibinfo  {journal} {Physical review letters}\
  }\textbf {\bibinfo {volume} {116}},\ \bibinfo {pages} {015902} (\bibinfo
  {year} {2016})}\BibitemShut {NoStop}%
\bibitem [{\citenamefont {Kurchan}\ \emph {et~al.}(2016)\citenamefont
  {Kurchan}, \citenamefont {Maimbourg},\ and\ \citenamefont
  {Zamponi}}]{kurchan2016statics}%
  \BibitemOpen
  \bibfield  {author} {\bibinfo {author} {\bibfnamefont {Jorge}\ \bibnamefont
  {Kurchan}}, \bibinfo {author} {\bibfnamefont {Thibaud}\ \bibnamefont
  {Maimbourg}}, \ and\ \bibinfo {author} {\bibfnamefont {Francesco}\
  \bibnamefont {Zamponi}},\ }\bibfield  {title} {\enquote {\bibinfo {title}
  {Statics and dynamics of infinite-dimensional liquids and glasses: a parallel
  and compact derivation},}\ }\href@noop {} {\bibfield  {journal} {\bibinfo
  {journal} {Journal of Statistical Mechanics: Theory and Experiment}\ }\textbf
  {\bibinfo {volume} {2016}},\ \bibinfo {pages} {033210} (\bibinfo {year}
  {2016})}\BibitemShut {NoStop}%
\bibitem [{\citenamefont {Agoritsas}\ \emph
  {et~al.}(2019{\natexlab{a}})\citenamefont {Agoritsas}, \citenamefont
  {Maimbourg},\ and\ \citenamefont {Zamponi}}]{agoritsas2019outA}%
  \BibitemOpen
  \bibfield  {author} {\bibinfo {author} {\bibfnamefont {Elisabeth}\
  \bibnamefont {Agoritsas}}, \bibinfo {author} {\bibfnamefont {Thibaud}\
  \bibnamefont {Maimbourg}}, \ and\ \bibinfo {author} {\bibfnamefont
  {Francesco}\ \bibnamefont {Zamponi}},\ }\bibfield  {title} {\enquote
  {\bibinfo {title} {Out-of-equilibrium dynamical equations of
  infinite-dimensional particle systems i. the isotropic case},}\ }\href@noop
  {} {\bibfield  {journal} {\bibinfo  {journal} {Journal of Physics A:
  Mathematical and Theoretical}\ }\textbf {\bibinfo {volume} {52}},\ \bibinfo
  {pages} {144002} (\bibinfo {year} {2019}{\natexlab{a}})}\BibitemShut
  {NoStop}%
\bibitem [{\citenamefont {Agoritsas}\ \emph
  {et~al.}(2019{\natexlab{b}})\citenamefont {Agoritsas}, \citenamefont
  {Maimbourg},\ and\ \citenamefont {Zamponi}}]{agoritsas2019outB}%
  \BibitemOpen
  \bibfield  {author} {\bibinfo {author} {\bibfnamefont {Elisabeth}\
  \bibnamefont {Agoritsas}}, \bibinfo {author} {\bibfnamefont {Thibaud}\
  \bibnamefont {Maimbourg}}, \ and\ \bibinfo {author} {\bibfnamefont
  {Francesco}\ \bibnamefont {Zamponi}},\ }\bibfield  {title} {\enquote
  {\bibinfo {title} {Out-of-equilibrium dynamical equations of
  infinite-dimensional particle systems. ii. the anisotropic case under shear
  strain},}\ }\href@noop {} {\bibfield  {journal} {\bibinfo  {journal} {Journal
  of Physics A: Mathematical and Theoretical}\ }\textbf {\bibinfo {volume}
  {52}},\ \bibinfo {pages} {334001} (\bibinfo {year}
  {2019}{\natexlab{b}})}\BibitemShut {NoStop}%
\bibitem [{\citenamefont {Kranendonk}\ and\ \citenamefont
  {Frenkel}(1991)}]{kranendonk1991computer}%
  \BibitemOpen
  \bibfield  {author} {\bibinfo {author} {\bibfnamefont {WGT}\ \bibnamefont
  {Kranendonk}}\ and\ \bibinfo {author} {\bibfnamefont {D}~\bibnamefont
  {Frenkel}},\ }\bibfield  {title} {\enquote {\bibinfo {title} {Computer
  simulation of solid-liquid coexistence in binary hard sphere mixtures},}\
  }\href@noop {} {\bibfield  {journal} {\bibinfo  {journal} {Molecular
  physics}\ }\textbf {\bibinfo {volume} {72}},\ \bibinfo {pages} {679--697}
  (\bibinfo {year} {1991})}\BibitemShut {NoStop}%
\bibitem [{\citenamefont {Grigera}\ and\ \citenamefont
  {Parisi}(2001)}]{grigera2001fast}%
  \BibitemOpen
  \bibfield  {author} {\bibinfo {author} {\bibfnamefont {Tom{\'a}s~S}\
  \bibnamefont {Grigera}}\ and\ \bibinfo {author} {\bibfnamefont {Giorgio}\
  \bibnamefont {Parisi}},\ }\bibfield  {title} {\enquote {\bibinfo {title}
  {Fast monte carlo algorithm for supercooled soft spheres},}\ }\href@noop {}
  {\bibfield  {journal} {\bibinfo  {journal} {Physical Review E}\ }\textbf
  {\bibinfo {volume} {63}},\ \bibinfo {pages} {045102} (\bibinfo {year}
  {2001})}\BibitemShut {NoStop}%
\bibitem [{\citenamefont {Berthier}\ \emph {et~al.}(2016)\citenamefont
  {Berthier}, \citenamefont {Charbonneau}, \citenamefont {Jin}, \citenamefont
  {Parisi}, \citenamefont {Seoane},\ and\ \citenamefont
  {Zamponi}}]{berthier2016growing}%
  \BibitemOpen
  \bibfield  {author} {\bibinfo {author} {\bibfnamefont {Ludovic}\ \bibnamefont
  {Berthier}}, \bibinfo {author} {\bibfnamefont {Patrick}\ \bibnamefont
  {Charbonneau}}, \bibinfo {author} {\bibfnamefont {Yuliang}\ \bibnamefont
  {Jin}}, \bibinfo {author} {\bibfnamefont {Giorgio}\ \bibnamefont {Parisi}},
  \bibinfo {author} {\bibfnamefont {Beatriz}\ \bibnamefont {Seoane}}, \ and\
  \bibinfo {author} {\bibfnamefont {Francesco}\ \bibnamefont {Zamponi}},\
  }\bibfield  {title} {\enquote {\bibinfo {title} {Growing timescales and
  lengthscales characterizing vibrations of amorphous solids},}\ }\href@noop {}
  {\bibfield  {journal} {\bibinfo  {journal} {Proceedings of the National
  Academy of Sciences}\ }\textbf {\bibinfo {volume} {113}},\ \bibinfo {pages}
  {8397--8401} (\bibinfo {year} {2016})}\BibitemShut {NoStop}%
\bibitem [{\citenamefont {Lubachevsky}\ and\ \citenamefont
  {Stillinger}(1990)}]{lubachevsky1990geometric}%
  \BibitemOpen
  \bibfield  {author} {\bibinfo {author} {\bibfnamefont {Boris~D}\ \bibnamefont
  {Lubachevsky}}\ and\ \bibinfo {author} {\bibfnamefont {Frank~H}\ \bibnamefont
  {Stillinger}},\ }\bibfield  {title} {\enquote {\bibinfo {title} {Geometric
  properties of random disk packings},}\ }\href@noop {} {\bibfield  {journal}
  {\bibinfo  {journal} {Journal of statistical Physics}\ }\textbf {\bibinfo
  {volume} {60}},\ \bibinfo {pages} {561--583} (\bibinfo {year}
  {1990})}\BibitemShut {NoStop}%
\bibitem [{\citenamefont {Goldstein}(2010)}]{Goldstein2010}%
  \BibitemOpen
  \bibfield  {author} {\bibinfo {author} {\bibfnamefont {Martin}\ \bibnamefont
  {Goldstein}},\ }\bibfield  {title} {\enquote {\bibinfo {title}
  {Communications: Comparison of activation barriers for the
  {J}ohari--{G}oldstein and alpha relaxations and its implications},}\ }\href
  {\doibase http://dx.doi.org/10.1063/1.3306562} {\bibfield  {journal}
  {\bibinfo  {journal} {J. Chem. Phys.}\ }\textbf {\bibinfo {volume} {132}},\
  \bibinfo {eid} {041104} (\bibinfo {year} {2010})}\BibitemShut {NoStop}%
\bibitem [{\citenamefont {Boublik}(1970)}]{B70}%
  \BibitemOpen
  \bibfield  {author} {\bibinfo {author} {\bibfnamefont {T.}~\bibnamefont
  {Boublik}},\ }\bibfield  {title} {\enquote {\bibinfo {title} {Hard sphere
  equation of state},}\ }\href@noop {} {\bibfield  {journal} {\bibinfo
  {journal} {J. Chem. Phys.}\ }\textbf {\bibinfo {volume} {53}},\ \bibinfo
  {pages} {471} (\bibinfo {year} {1970})}\BibitemShut {NoStop}%
\bibitem [{\citenamefont {Charbonneau}\ \emph
  {et~al.}(2014{\natexlab{a}})\citenamefont {Charbonneau}, \citenamefont {Jin},
  \citenamefont {Parisi},\ and\ \citenamefont {Zamponi}}]{Charbonneau2014}%
  \BibitemOpen
  \bibfield  {author} {\bibinfo {author} {\bibfnamefont {Patrick}\ \bibnamefont
  {Charbonneau}}, \bibinfo {author} {\bibfnamefont {Yuliang}\ \bibnamefont
  {Jin}}, \bibinfo {author} {\bibfnamefont {Giorgio}\ \bibnamefont {Parisi}}, \
  and\ \bibinfo {author} {\bibfnamefont {Francesco}\ \bibnamefont {Zamponi}},\
  }\bibfield  {title} {\enquote {\bibinfo {title} {Hopping and the
  {S}tokes--{E}instein relation breakdown in simple glass formers},}\ }\href
  {\doibase 10.1073/pnas.1417182111} {\bibfield  {journal} {\bibinfo  {journal}
  {Proc. Natl. Acad. Sci. U.S.A.}\ }\textbf {\bibinfo {volume} {111}},\
  \bibinfo {pages} {15025--15030} (\bibinfo {year}
  {2014}{\natexlab{a}})}\BibitemShut {NoStop}%
\bibitem [{\citenamefont {Donev}\ \emph {et~al.}(2005)\citenamefont {Donev},
  \citenamefont {Torquato},\ and\ \citenamefont {Stillinger}}]{donev2005pair}%
  \BibitemOpen
  \bibfield  {author} {\bibinfo {author} {\bibfnamefont {Aleksandar}\
  \bibnamefont {Donev}}, \bibinfo {author} {\bibfnamefont {Salvatore}\
  \bibnamefont {Torquato}}, \ and\ \bibinfo {author} {\bibfnamefont {Frank~H}\
  \bibnamefont {Stillinger}},\ }\bibfield  {title} {\enquote {\bibinfo {title}
  {Pair correlation function characteristics of nearly jammed disordered and
  ordered hard-sphere packings},}\ }\href@noop {} {\bibfield  {journal}
  {\bibinfo  {journal} {Physical Review E}\ }\textbf {\bibinfo {volume} {71}},\
  \bibinfo {pages} {011105} (\bibinfo {year} {2005})}\BibitemShut {NoStop}%
\bibitem [{\citenamefont {Bouchaud}\ \emph {et~al.}(1998)\citenamefont
  {Bouchaud}, \citenamefont {Cugliandolo}, \citenamefont {Kurchan},\ and\
  \citenamefont {M{\'e}zard}}]{bouchaud1998out}%
  \BibitemOpen
  \bibfield  {author} {\bibinfo {author} {\bibfnamefont {Jean-Philippe}\
  \bibnamefont {Bouchaud}}, \bibinfo {author} {\bibfnamefont {Leticia~F}\
  \bibnamefont {Cugliandolo}}, \bibinfo {author} {\bibfnamefont {Jorge}\
  \bibnamefont {Kurchan}}, \ and\ \bibinfo {author} {\bibfnamefont {Marc}\
  \bibnamefont {M{\'e}zard}},\ }\bibfield  {title} {\enquote {\bibinfo {title}
  {Out of equilibrium dynamics in spin-glasses and other glassy systems},}\
  }\href@noop {} {\bibfield  {journal} {\bibinfo  {journal} {Spin glasses and
  random fields}\ }\textbf {\bibinfo {volume} {12}},\ \bibinfo {pages} {161}
  (\bibinfo {year} {1998})}\BibitemShut {NoStop}%
\bibitem [{\citenamefont {Lees}\ and\ \citenamefont
  {Edwards}(1972)}]{lees1972computer}%
  \BibitemOpen
  \bibfield  {author} {\bibinfo {author} {\bibfnamefont {AW}~\bibnamefont
  {Lees}}\ and\ \bibinfo {author} {\bibfnamefont {SF}~\bibnamefont {Edwards}},\
  }\bibfield  {title} {\enquote {\bibinfo {title} {The computer study of
  transport processes under extreme conditions},}\ }\href@noop {} {\bibfield
  {journal} {\bibinfo  {journal} {J. Phys. Condens. Matter}\ }\textbf {\bibinfo
  {volume} {5}},\ \bibinfo {pages} {1921} (\bibinfo {year} {1972})}\BibitemShut
  {NoStop}%
\bibitem [{\citenamefont {Jin}\ and\ \citenamefont
  {Yoshino}(2017)}]{jin2017exploring}%
  \BibitemOpen
  \bibfield  {author} {\bibinfo {author} {\bibfnamefont {Yuliang}\ \bibnamefont
  {Jin}}\ and\ \bibinfo {author} {\bibfnamefont {Hajime}\ \bibnamefont
  {Yoshino}},\ }\bibfield  {title} {\enquote {\bibinfo {title} {Exploring the
  complex free-energy landscape of the simplest glass by rheology},}\
  }\href@noop {} {\bibfield  {journal} {\bibinfo  {journal} {Nature
  communications}\ }\textbf {\bibinfo {volume} {8}},\ \bibinfo {pages} {1--8}
  (\bibinfo {year} {2017})}\BibitemShut {NoStop}%
\bibitem [{\citenamefont {Charbonneau}\ \emph
  {et~al.}(2015{\natexlab{a}})\citenamefont {Charbonneau}, \citenamefont {Jin},
  \citenamefont {Parisi}, \citenamefont {Rainone}, \citenamefont {Seoane},\
  and\ \citenamefont {Zamponi}}]{charbonneau2015numerical}%
  \BibitemOpen
  \bibfield  {author} {\bibinfo {author} {\bibfnamefont {Patrick}\ \bibnamefont
  {Charbonneau}}, \bibinfo {author} {\bibfnamefont {Yuliang}\ \bibnamefont
  {Jin}}, \bibinfo {author} {\bibfnamefont {Giorgio}\ \bibnamefont {Parisi}},
  \bibinfo {author} {\bibfnamefont {Corrado}\ \bibnamefont {Rainone}}, \bibinfo
  {author} {\bibfnamefont {Beatriz}\ \bibnamefont {Seoane}}, \ and\ \bibinfo
  {author} {\bibfnamefont {Francesco}\ \bibnamefont {Zamponi}},\ }\bibfield
  {title} {\enquote {\bibinfo {title} {Numerical detection of the gardner
  transition in a mean-field glass former},}\ }\href@noop {} {\bibfield
  {journal} {\bibinfo  {journal} {Physical Review E}\ }\textbf {\bibinfo
  {volume} {92}},\ \bibinfo {pages} {012316} (\bibinfo {year}
  {2015}{\natexlab{a}})}\BibitemShut {NoStop}%
\bibitem [{\citenamefont {Seoane}\ and\ \citenamefont
  {Zamponi}(2018)}]{seoane2018spin}%
  \BibitemOpen
  \bibfield  {author} {\bibinfo {author} {\bibfnamefont {Beatriz}\ \bibnamefont
  {Seoane}}\ and\ \bibinfo {author} {\bibfnamefont {Francesco}\ \bibnamefont
  {Zamponi}},\ }\bibfield  {title} {\enquote {\bibinfo {title} {Spin-glass-like
  aging in colloidal and granular glasses},}\ }\href@noop {} {\bibfield
  {journal} {\bibinfo  {journal} {Soft Matter}\ }\textbf {\bibinfo {volume}
  {14}},\ \bibinfo {pages} {5222--5234} (\bibinfo {year} {2018})}\BibitemShut
  {NoStop}%
\bibitem [{\citenamefont {Charbonneau}\ \emph
  {et~al.}(2014{\natexlab{b}})\citenamefont {Charbonneau}, \citenamefont
  {Kurchan}, \citenamefont {Parisi}, \citenamefont {Urbani},\ and\
  \citenamefont {Zamponi}}]{charbonneau2014fractal}%
  \BibitemOpen
  \bibfield  {author} {\bibinfo {author} {\bibfnamefont {Patrick}\ \bibnamefont
  {Charbonneau}}, \bibinfo {author} {\bibfnamefont {Jorge}\ \bibnamefont
  {Kurchan}}, \bibinfo {author} {\bibfnamefont {Giorgio}\ \bibnamefont
  {Parisi}}, \bibinfo {author} {\bibfnamefont {Pierfrancesco}\ \bibnamefont
  {Urbani}}, \ and\ \bibinfo {author} {\bibfnamefont {Francesco}\ \bibnamefont
  {Zamponi}},\ }\bibfield  {title} {\enquote {\bibinfo {title} {Fractal free
  energy landscapes in structural glasses},}\ }\href@noop {} {\bibfield
  {journal} {\bibinfo  {journal} {Nature communications}\ }\textbf {\bibinfo
  {volume} {5}},\ \bibinfo {pages} {1--6} (\bibinfo {year}
  {2014}{\natexlab{b}})}\BibitemShut {NoStop}%
\bibitem [{\citenamefont {Rainone}\ \emph {et~al.}(2015)\citenamefont
  {Rainone}, \citenamefont {Urbani}, \citenamefont {Yoshino},\ and\
  \citenamefont {Zamponi}}]{rainone2015following}%
  \BibitemOpen
  \bibfield  {author} {\bibinfo {author} {\bibfnamefont {Corrado}\ \bibnamefont
  {Rainone}}, \bibinfo {author} {\bibfnamefont {Pierfrancesco}\ \bibnamefont
  {Urbani}}, \bibinfo {author} {\bibfnamefont {Hajime}\ \bibnamefont
  {Yoshino}}, \ and\ \bibinfo {author} {\bibfnamefont {Francesco}\ \bibnamefont
  {Zamponi}},\ }\bibfield  {title} {\enquote {\bibinfo {title} {Following the
  evolution of hard sphere glasses in infinite dimensions under external
  perturbations: Compression and shear strain},}\ }\href@noop {} {\bibfield
  {journal} {\bibinfo  {journal} {Physical review letters}\ }\textbf {\bibinfo
  {volume} {114}},\ \bibinfo {pages} {015701} (\bibinfo {year}
  {2015})}\BibitemShut {NoStop}%
\bibitem [{\citenamefont {Jin}\ \emph {et~al.}(2018)\citenamefont {Jin},
  \citenamefont {Urbani}, \citenamefont {Zamponi},\ and\ \citenamefont
  {Yoshino}}]{jin2018stability}%
  \BibitemOpen
  \bibfield  {author} {\bibinfo {author} {\bibfnamefont {Yuliang}\ \bibnamefont
  {Jin}}, \bibinfo {author} {\bibfnamefont {Pierfrancesco}\ \bibnamefont
  {Urbani}}, \bibinfo {author} {\bibfnamefont {Francesco}\ \bibnamefont
  {Zamponi}}, \ and\ \bibinfo {author} {\bibfnamefont {Hajime}\ \bibnamefont
  {Yoshino}},\ }\bibfield  {title} {\enquote {\bibinfo {title} {A
  stability-reversibility map unifies elasticity, plasticity, yielding, and
  jamming in hard sphere glasses},}\ }\href@noop {} {\bibfield  {journal}
  {\bibinfo  {journal} {Science advances}\ }\textbf {\bibinfo {volume} {4}},\
  \bibinfo {pages} {eaat6387} (\bibinfo {year} {2018})}\BibitemShut {NoStop}%
\bibitem [{\citenamefont {Altieri}\ and\ \citenamefont
  {Zamponi}(2019)}]{altieri2019mean}%
  \BibitemOpen
  \bibfield  {author} {\bibinfo {author} {\bibfnamefont {Ada}\ \bibnamefont
  {Altieri}}\ and\ \bibinfo {author} {\bibfnamefont {Francesco}\ \bibnamefont
  {Zamponi}},\ }\bibfield  {title} {\enquote {\bibinfo {title} {Mean-field
  stability map of hard-sphere glasses},}\ }\href@noop {} {\bibfield  {journal}
  {\bibinfo  {journal} {Physical Review E}\ }\textbf {\bibinfo {volume}
  {100}},\ \bibinfo {pages} {032140} (\bibinfo {year} {2019})}\BibitemShut
  {NoStop}%
\bibitem [{\citenamefont {Liao}\ and\ \citenamefont
  {Berthier}(2019)}]{liao2019hierarchical}%
  \BibitemOpen
  \bibfield  {author} {\bibinfo {author} {\bibfnamefont {Qinyi}\ \bibnamefont
  {Liao}}\ and\ \bibinfo {author} {\bibfnamefont {Ludovic}\ \bibnamefont
  {Berthier}},\ }\bibfield  {title} {\enquote {\bibinfo {title} {Hierarchical
  landscape of hard disk glasses},}\ }\href@noop {} {\bibfield  {journal}
  {\bibinfo  {journal} {Physical Review X}\ }\textbf {\bibinfo {volume} {9}},\
  \bibinfo {pages} {011049} (\bibinfo {year} {2019})}\BibitemShut {NoStop}%
\bibitem [{\citenamefont {Charbonneau}\ and\ \citenamefont
  {Yaida}(2017)}]{charbonneau2017nontrivial}%
  \BibitemOpen
  \bibfield  {author} {\bibinfo {author} {\bibfnamefont {Patrick}\ \bibnamefont
  {Charbonneau}}\ and\ \bibinfo {author} {\bibfnamefont {Sho}\ \bibnamefont
  {Yaida}},\ }\bibfield  {title} {\enquote {\bibinfo {title} {Nontrivial
  critical fixed point for replica-symmetry-breaking transitions},}\
  }\href@noop {} {\bibfield  {journal} {\bibinfo  {journal} {Physical review
  letters}\ }\textbf {\bibinfo {volume} {118}},\ \bibinfo {pages} {215701}
  (\bibinfo {year} {2017})}\BibitemShut {NoStop}%
\bibitem [{\citenamefont {Li}\ \emph {et~al.}(2021)\citenamefont {Li},
  \citenamefont {Jin}, \citenamefont {Jiang},\ and\ \citenamefont
  {Chen}}]{li2021determining}%
  \BibitemOpen
  \bibfield  {author} {\bibinfo {author} {\bibfnamefont {Huaping}\ \bibnamefont
  {Li}}, \bibinfo {author} {\bibfnamefont {Yuliang}\ \bibnamefont {Jin}},
  \bibinfo {author} {\bibfnamefont {Ying}\ \bibnamefont {Jiang}}, \ and\
  \bibinfo {author} {\bibfnamefont {Jeff~ZY}\ \bibnamefont {Chen}},\ }\bibfield
   {title} {\enquote {\bibinfo {title} {Determining the nonequilibrium
  criticality of a gardner transition via a hybrid study of molecular
  simulations and machine learning},}\ }\href@noop {} {\bibfield  {journal}
  {\bibinfo  {journal} {Proceedings of the National Academy of Sciences}\
  }\textbf {\bibinfo {volume} {118}} (\bibinfo {year} {2021})}\BibitemShut
  {NoStop}%
\bibitem [{\citenamefont {Franz}\ \emph {et~al.}(2015)\citenamefont {Franz},
  \citenamefont {Parisi}, \citenamefont {Urbani},\ and\ \citenamefont
  {Zamponi}}]{franz2015universal}%
  \BibitemOpen
  \bibfield  {author} {\bibinfo {author} {\bibfnamefont {Silvio}\ \bibnamefont
  {Franz}}, \bibinfo {author} {\bibfnamefont {Giorgio}\ \bibnamefont {Parisi}},
  \bibinfo {author} {\bibfnamefont {Pierfrancesco}\ \bibnamefont {Urbani}}, \
  and\ \bibinfo {author} {\bibfnamefont {Francesco}\ \bibnamefont {Zamponi}},\
  }\bibfield  {title} {\enquote {\bibinfo {title} {Universal spectrum of normal
  modes in low-temperature glasses},}\ }\href@noop {} {\bibfield  {journal}
  {\bibinfo  {journal} {Proceedings of the National Academy of Sciences}\
  }\textbf {\bibinfo {volume} {112}},\ \bibinfo {pages} {14539--14544}
  (\bibinfo {year} {2015})}\BibitemShut {NoStop}%
\bibitem [{\citenamefont {Charbonneau}\ \emph
  {et~al.}(2015{\natexlab{b}})\citenamefont {Charbonneau}, \citenamefont
  {Corwin}, \citenamefont {Parisi},\ and\ \citenamefont
  {Zamponi}}]{charbonneau2015jamming}%
  \BibitemOpen
  \bibfield  {author} {\bibinfo {author} {\bibfnamefont {Patrick}\ \bibnamefont
  {Charbonneau}}, \bibinfo {author} {\bibfnamefont {Eric~I}\ \bibnamefont
  {Corwin}}, \bibinfo {author} {\bibfnamefont {Giorgio}\ \bibnamefont
  {Parisi}}, \ and\ \bibinfo {author} {\bibfnamefont {Francesco}\ \bibnamefont
  {Zamponi}},\ }\bibfield  {title} {\enquote {\bibinfo {title} {Jamming
  criticality revealed by removing localized buckling excitations},}\
  }\href@noop {} {\bibfield  {journal} {\bibinfo  {journal} {Physical review
  letters}\ }\textbf {\bibinfo {volume} {114}},\ \bibinfo {pages} {125504}
  (\bibinfo {year} {2015}{\natexlab{b}})}\BibitemShut {NoStop}%
\bibitem [{\citenamefont {Charbonneau}\ \emph {et~al.}(2016)\citenamefont
  {Charbonneau}, \citenamefont {Corwin}, \citenamefont {Parisi}, \citenamefont
  {Poncet},\ and\ \citenamefont {Zamponi}}]{charbonneau2016universal}%
  \BibitemOpen
  \bibfield  {author} {\bibinfo {author} {\bibfnamefont {Patrick}\ \bibnamefont
  {Charbonneau}}, \bibinfo {author} {\bibfnamefont {Eric~I}\ \bibnamefont
  {Corwin}}, \bibinfo {author} {\bibfnamefont {Giorgio}\ \bibnamefont
  {Parisi}}, \bibinfo {author} {\bibfnamefont {Alexis}\ \bibnamefont {Poncet}},
  \ and\ \bibinfo {author} {\bibfnamefont {Francesco}\ \bibnamefont
  {Zamponi}},\ }\bibfield  {title} {\enquote {\bibinfo {title} {Universal
  non-debye scaling in the density of states of amorphous solids},}\
  }\href@noop {} {\bibfield  {journal} {\bibinfo  {journal} {Physical review
  letters}\ }\textbf {\bibinfo {volume} {117}},\ \bibinfo {pages} {045503}
  (\bibinfo {year} {2016})}\BibitemShut {NoStop}%
\bibitem [{\citenamefont {Dennis}\ and\ \citenamefont
  {Corwin}(2020)}]{dennis2020jamming}%
  \BibitemOpen
  \bibfield  {author} {\bibinfo {author} {\bibfnamefont {RC}~\bibnamefont
  {Dennis}}\ and\ \bibinfo {author} {\bibfnamefont {EI}~\bibnamefont
  {Corwin}},\ }\bibfield  {title} {\enquote {\bibinfo {title} {Jamming energy
  landscape is hierarchical and ultrametric},}\ }\href@noop {} {\bibfield
  {journal} {\bibinfo  {journal} {Physical review letters}\ }\textbf {\bibinfo
  {volume} {124}},\ \bibinfo {pages} {078002} (\bibinfo {year}
  {2020})}\BibitemShut {NoStop}%
\bibitem [{\citenamefont {Scalliet}\ \emph {et~al.}(2019)\citenamefont
  {Scalliet}, \citenamefont {Berthier},\ and\ \citenamefont
  {Zamponi}}]{scalliet2019marginally}%
  \BibitemOpen
  \bibfield  {author} {\bibinfo {author} {\bibfnamefont {Camille}\ \bibnamefont
  {Scalliet}}, \bibinfo {author} {\bibfnamefont {Ludovic}\ \bibnamefont
  {Berthier}}, \ and\ \bibinfo {author} {\bibfnamefont {Francesco}\
  \bibnamefont {Zamponi}},\ }\bibfield  {title} {\enquote {\bibinfo {title}
  {Marginally stable phases in mean-field structural glasses},}\ }\href@noop {}
  {\bibfield  {journal} {\bibinfo  {journal} {Physical Review E}\ }\textbf
  {\bibinfo {volume} {99}},\ \bibinfo {pages} {012107} (\bibinfo {year}
  {2019})}\BibitemShut {NoStop}%
\bibitem [{\citenamefont {Scalliet}\ and\ \citenamefont
  {Berthier}(2019)}]{scalliet2019rejuvenation}%
  \BibitemOpen
  \bibfield  {author} {\bibinfo {author} {\bibfnamefont {Camille}\ \bibnamefont
  {Scalliet}}\ and\ \bibinfo {author} {\bibfnamefont {Ludovic}\ \bibnamefont
  {Berthier}},\ }\bibfield  {title} {\enquote {\bibinfo {title} {Rejuvenation
  and memory effects in a structural glass},}\ }\href@noop {} {\bibfield
  {journal} {\bibinfo  {journal} {Physical review letters}\ }\textbf {\bibinfo
  {volume} {122}},\ \bibinfo {pages} {255502} (\bibinfo {year}
  {2019})}\BibitemShut {NoStop}%
\bibitem [{\citenamefont {Scalliet}\ \emph {et~al.}(2017)\citenamefont
  {Scalliet}, \citenamefont {Berthier},\ and\ \citenamefont
  {Zamponi}}]{scalliet2017absence}%
  \BibitemOpen
  \bibfield  {author} {\bibinfo {author} {\bibfnamefont {Camille}\ \bibnamefont
  {Scalliet}}, \bibinfo {author} {\bibfnamefont {Ludovic}\ \bibnamefont
  {Berthier}}, \ and\ \bibinfo {author} {\bibfnamefont {Francesco}\
  \bibnamefont {Zamponi}},\ }\bibfield  {title} {\enquote {\bibinfo {title}
  {Absence of marginal stability in a structural glass},}\ }\href@noop {}
  {\bibfield  {journal} {\bibinfo  {journal} {Physical review letters}\
  }\textbf {\bibinfo {volume} {119}},\ \bibinfo {pages} {205501} (\bibinfo
  {year} {2017})}\BibitemShut {NoStop}%
\bibitem [{\citenamefont {Seguin}\ and\ \citenamefont
  {Dauchot}(2016)}]{seguin2016experimental}%
  \BibitemOpen
  \bibfield  {author} {\bibinfo {author} {\bibfnamefont {Antoine}\ \bibnamefont
  {Seguin}}\ and\ \bibinfo {author} {\bibfnamefont {Olivier}\ \bibnamefont
  {Dauchot}},\ }\bibfield  {title} {\enquote {\bibinfo {title} {Experimental
  evidence of the gardner phase in a granular glass},}\ }\href@noop {}
  {\bibfield  {journal} {\bibinfo  {journal} {Physical review letters}\
  }\textbf {\bibinfo {volume} {117}},\ \bibinfo {pages} {228001} (\bibinfo
  {year} {2016})}\BibitemShut {NoStop}%
\bibitem [{\citenamefont {Hammond}\ and\ \citenamefont
  {Corwin}(2020)}]{hammond2020experimental}%
  \BibitemOpen
  \bibfield  {author} {\bibinfo {author} {\bibfnamefont {Andrew~P}\
  \bibnamefont {Hammond}}\ and\ \bibinfo {author} {\bibfnamefont {Eric~I}\
  \bibnamefont {Corwin}},\ }\bibfield  {title} {\enquote {\bibinfo {title}
  {Experimental observation of the marginal glass phase in a colloidal
  glass},}\ }\href@noop {} {\bibfield  {journal} {\bibinfo  {journal}
  {Proceedings of the National Academy of Sciences}\ }\textbf {\bibinfo
  {volume} {117}},\ \bibinfo {pages} {5714--5718} (\bibinfo {year}
  {2020})}\BibitemShut {NoStop}%
\bibitem [{\citenamefont {Zargar}\ \emph {et~al.}(2017)\citenamefont {Zargar},
  \citenamefont {DeGiuli},\ and\ \citenamefont {Bonn}}]{zargar2017scaling}%
  \BibitemOpen
  \bibfield  {author} {\bibinfo {author} {\bibfnamefont {Rojman}\ \bibnamefont
  {Zargar}}, \bibinfo {author} {\bibfnamefont {Eric}\ \bibnamefont {DeGiuli}},
  \ and\ \bibinfo {author} {\bibfnamefont {Daniel}\ \bibnamefont {Bonn}},\
  }\bibfield  {title} {\enquote {\bibinfo {title} {Scaling for hard-sphere
  colloidal glasses near jamming},}\ }\href@noop {} {\bibfield  {journal}
  {\bibinfo  {journal} {EPL (Europhysics Letters)}\ }\textbf {\bibinfo {volume}
  {116}},\ \bibinfo {pages} {68004} (\bibinfo {year} {2017})}\BibitemShut
  {NoStop}%
\bibitem [{\citenamefont {Wang}\ \emph {et~al.}(2022)\citenamefont {Wang},
  \citenamefont {Shang}, \citenamefont {Jin},\ and\ \citenamefont
  {Zhang}}]{doi:10.1073/pnas.2204879119}%
  \BibitemOpen
  \bibfield  {author} {\bibinfo {author} {\bibfnamefont {Yinqiao}\ \bibnamefont
  {Wang}}, \bibinfo {author} {\bibfnamefont {Jin}\ \bibnamefont {Shang}},
  \bibinfo {author} {\bibfnamefont {Yuliang}\ \bibnamefont {Jin}}, \ and\
  \bibinfo {author} {\bibfnamefont {Jie}\ \bibnamefont {Zhang}},\ }\bibfield
  {title} {\enquote {\bibinfo {title} {Experimental observations of marginal
  criticality in granular materials},}\ }\href@noop {} {\bibfield  {journal}
  {\bibinfo  {journal} {Proceedings of the National Academy of Sciences}\
  }\textbf {\bibinfo {volume} {119}},\ \bibinfo {pages} {e2204879119} (\bibinfo
  {year} {2022})}\BibitemShut {NoStop}%
\bibitem [{\citenamefont {Kool}\ \emph {et~al.}(2022)\citenamefont {Kool},
  \citenamefont {Charbonneau},\ and\ \citenamefont
  {Daniels}}]{kool2022gardner}%
  \BibitemOpen
  \bibfield  {author} {\bibinfo {author} {\bibfnamefont {Lars}\ \bibnamefont
  {Kool}}, \bibinfo {author} {\bibfnamefont {Patrick}\ \bibnamefont
  {Charbonneau}}, \ and\ \bibinfo {author} {\bibfnamefont {Karen~E}\
  \bibnamefont {Daniels}},\ }\bibfield  {title} {\enquote {\bibinfo {title}
  {Gardner-like transition from variable to persistent force contacts in
  granular crystals},}\ }\href@noop {} {\bibfield  {journal} {\bibinfo
  {journal} {arXiv preprint arXiv:2205.06794}\ } (\bibinfo {year}
  {2022})}\BibitemShut {NoStop}%
\bibitem [{\citenamefont {Charbonneau}\ \emph {et~al.}(2019)\citenamefont
  {Charbonneau}, \citenamefont {Corwin}, \citenamefont {Fu}, \citenamefont
  {Tsekenis},\ and\ \citenamefont {van Der~Naald}}]{charbonneau2019glassy}%
  \BibitemOpen
  \bibfield  {author} {\bibinfo {author} {\bibfnamefont {Patrick}\ \bibnamefont
  {Charbonneau}}, \bibinfo {author} {\bibfnamefont {Eric~I}\ \bibnamefont
  {Corwin}}, \bibinfo {author} {\bibfnamefont {Lin}\ \bibnamefont {Fu}},
  \bibinfo {author} {\bibfnamefont {Georgios}\ \bibnamefont {Tsekenis}}, \ and\
  \bibinfo {author} {\bibfnamefont {Michael}\ \bibnamefont {van Der~Naald}},\
  }\bibfield  {title} {\enquote {\bibinfo {title} {Glassy, gardner-like
  phenomenology in minimally polydisperse crystalline systems},}\ }\href@noop
  {} {\bibfield  {journal} {\bibinfo  {journal} {Physical Review E}\ }\textbf
  {\bibinfo {volume} {99}},\ \bibinfo {pages} {020901} (\bibinfo {year}
  {2019})}\BibitemShut {NoStop}%
\end{thebibliography}%

\end{document}